\documentclass[smallextended]{svjour3} 

\usepackage[utf8]{inputenc}
\usepackage[T1]{fontenc}

\usepackage{array}
\usepackage{subcaption}
\usepackage{graphicx}
\usepackage{multirow}
\graphicspath{ {figs/} }
\usepackage{balance}
\usepackage[linesnumbered,ruled,vlined]{algorithm2e}
\let\oldnl\nl% Store \nl in \oldnl
\newcommand{\nonl}{\renewcommand{\nl}{\let\nl\oldnl}}% Remove line number for one line
\usepackage{algorithmic}
\usepackage{esvect}
\usepackage{booktabs}
\usepackage{longtable}
\usepackage{tabularx}
\usepackage{anyfontsize}
\usepackage[all]{nowidow}
\usepackage[normalem]{ulem}
\usepackage{xcolor}
\usepackage{colortbl}
\usepackage{enumitem}
\usepackage{makecell}
\usepackage{lineno}
\usepackage{natbib}

\newcolumntype{Y}{>{\centering\arraybackslash}X}

\definecolor{commentGray}{RGB}{120,120,120}
\renewcommand{\algorithmiccomment}[1]{\bgroup\color{commentGray}{//#1}\egroup}

\usepackage{framed}
\usepackage{tikz}
\definecolor{light-gray}{gray}{0.9}
\usepackage[framemethod=TikZ]{mdframed}
\mdfsetup{skipabove=5pt,skipbelow=3pt}
\usepackage{lipsum}
\mdfdefinestyle{RQFrame}{%
	linecolor=black,
	outerlinewidth=0.15pt,
	roundcorner=3pt,
	innertopmargin=2pt,
	innerbottommargin=2pt,
	innerrightmargin=4pt,
	innerleftmargin=4pt,
	backgroundcolor=light-gray}
\definecolor{javagreen}{rgb}{0.25,0.5,0.35} % comments

\newcommand{\Odif}{{\text{\fontsize{8}{8}$\textsf{MTEC } $}}}
\newcommand{\Ofit}{{\text{\fontsize{8}{8}$\textsf{STEC } $}}}
\newcommand{\Dfit}{{\text{{\fontsize{8}{8}$D_{\textsf{STEC }} $}}}}
\newcommand{\Ddif}{{\text{{\fontsize{8}{8}$D_{\textsf{MTEC }} $}}}}

\newcommand{\Didif}[1]{{\text{{\fontsize{8}{8}$D^{#1}_{\textsf{MTEC }} $}}}}

\newcounter{commentnumber}
\usepackage[hyphens]{url}
\usepackage[hidelinks]{hyperref}
\hypersetup{breaklinks=true}
\usepackage{amsthm}
\usepackage{newtxmath} % for Latex 2020

\journalname{Empirical Software Engineering}

\begin{document}

\title{Evaluating the Impact of Flaky Simulators on Testing Autonomous Driving Systems}
\titlerunning{Evaluating the Impact of Flaky Simulators}

\author{Mohammad Hossein Amini \and Shervin Naseri \and Shiva Nejati}
\institute{Mohammad Hossein Amini \at {University of Ottawa, Canada}\\
\email{mh.amini@uottawa.ca}\\
\and
Shervin Naseri \at {University of Ottawa, Canada}\\
\email{snase041@uottawa.ca}\\
\and
Shiva Nejati \at {University of Ottawa, Canada}\\
\email{snejati@uottawa.ca}}

% \twocolumn
%\input{letterMajorReview.tex}
%\input{answerMajorReview.tex}

%\input{letterMinorReview.tex}
%\input{answerMinorRevision.tex}

\date{Received: date / Accepted: date}

\maketitle
\thispagestyle{empty}

\begin{abstract}
%This paper presents the first comprehensive study on test flakiness for simulation-based testing of autonomous driving systems (ADS). 

Simulators are widely used to test Autonomous Driving Systems (ADS), but their potential flakiness can lead to inconsistent test results. We investigate test flakiness in simulation-based testing of ADS by addressing two key questions: (1)~How do flaky ADS simulations impact automated testing that relies on randomized algorithms? and (2)~Can machine learning (ML)  effectively identify flaky ADS tests while decreasing the required number of test reruns?  Our empirical results, obtained from two widely-used open-source ADS simulators and five diverse ADS test setups, show that  test flakiness in ADS is a common occurrence and can significantly impact the test results obtained by randomized algorithms. Further, our ML classifiers  effectively identify flaky ADS tests  using only  a single test run,  achieving F1-scores of $85$\%, $82$\% and $96$\% for three different ADS test setups. Our classifiers significantly outperform our non-ML baseline, which requires executing tests at least twice, by $31$\%, $21$\%, and $13$\% in F1-score performance, respectively. We conclude with a discussion on the scope, implications and limitations of our study. 
% \linelabel{R2C3L1}
We provide our complete replication package in a Github repository \citep{github}.
\end{abstract}


%ML classifiers effectively identify flaky ADS tests with one test run, achieving F1-scores of 82%, 85%, and 96% for three setups. With four or fewer reruns, F1-scores increase to 94%, 89%, and 97%, surpassing non-ML baselines by 15%, 7%, and 2%.

%Notably, ML classifiers successfully identify flaky ADS tests, requiring a maximum of four re-executions per test. These classifiers achieve a recall of at least 76% and a precision of at least 89%, substantially surpassing a non-ML benchmark.

%Simulators, however, exhibit flaky behaviour due to a variety of factors such as concurrency, numerical instability, stochastic elements, and environment uncertainty (different intial conditions).

% !TEX root =  ../main.tex
\section{Introduction}
\label{sec:intro}
Simulation-based  testing is the technique of choice for at-scale verification of systems with high levels of autonomy, e.g., autonomous driving systems (ADS)~\citep{RaquelCompany,ahlgren2021facebook,raja2018,BorgANJS21}. Simulators exercise very large numbers of system-usage scenarios that would be prohibitively expensive or time-consuming to enact in the real world.  Many ADS rely on deep neural networks (DNNs) either partially or entirely. Testing DNN-enabled ADS using simulators can reveal failures that would remain undetected when testing DNNs individually and without embedding them into a closed-loop simulation environment~\citep{DBLP:conf/icst/HaqSNB20,DBLP:journals/ese/HaqSNB21}. This highlights the significance of simulation-based ADS testing.
% This is because simulation-based testing reveals the cumulative effects of minor prediction errors over time, potentially leading to critical safety violations such as pedestrian collisions, which cannot be detected through isolated offline testing of the ADS component.

Recent studies have noted that ADS simulators can be flaky~\citep{salvo,arxivintro}, and an industry survey further emphasizes this flakiness as a major challenge to reproducibility in the field of robotic simulation~\citep{afsoon-robotics}. Current simulation-based ADS testing methods either discard flaky tests to avoid inaccuracies~\citep{arxivintro} or limit test scenario variables to reduce flakiness~\citep{salvo}. These recent studies on ADS testing highlight the significance of addressing flaky tests in virtual environments and simulators since flaky tests lower the reliability of simulation-based testing for safety-critical systems~\citep{arxivintro,salvo}. However, to the best of our knowledge, there is no systematic study on the prevalence, impact, and potential mitigation strategies for test flakiness in simulation-based ADS testing.

In contrast, test flakiness has been widely studied in software code, where flaky tests are those exhibiting non-deterministic behavior by passing or failing over different runs when applied to the same codebase~\citep{survey,DBLP:conf/sigsoft/LuoHEM14,DBLP:conf/icse/Alshammari0H021}. These tests can be problematic and time-consuming, as they make it difficult to determine whether a code modification has caused a test to fail or if the failure is due to the test's flakiness.

%Darko's paper FSE 2014 or around that time.
%Q. Luo, F. Hariri, L. Eloussi, and D. Marinov. 2014. An empirical analysis of flaky tests. In Proceedings of the Symposium on the Foundations of Software Engineering (FSE’14). 643–653.

%Simulation for Robotics Test Automation: Developer Perspectives
%They mention flaky tests in (https://arxiv.org/pdf/2212.04769.pdf). They say they had 1\% to 5\% flaky tests that they decided to discard. 

In this paper, we present a systematic study on flakiness in simulation-based ADS testing. Though the investigated questions are relevant to all autonomous system simulators that exhibit non-determinism, we focus on ADS simulators due to their importance and wide-spread use. Our study aims to ultimately answer the following   questions: \emph{RQ1. How do flaky ADS simulations impact automated testing that relies on randomized algorithms?} and \emph{RQ2. Can machine learning
(ML) effectively identify flaky ADS tests while decreasing the
required number of test reruns?} We answer these questions using two widely-used, open-source ADS simulators, CARLA~\citep{carlapaper} and BeamNG~\citep{beamng}, and based on five different ADS test setups. These test setups enable us to examine test flakiness across various ADS types while considering a diverse range of ADS input variables for testing purposes. These setups include (1)~CARLA with its builtin PID-based ADS~\citep{carlatm}; (2)~CARLA with Pylot, a modular DNN-enabled ADS~\citep{gog2021pylot,samota}; (3)~CARLA with Transfuser, an end-to-end DNN-enabled ADS~\citep{transfuser,haq2022manyobjective}, (4)~BeamNG with its AI-engine~\citep{beamng}, and (5)~the BeamNG test setup from the tool competition track of the SBFT workshop~\citep{sbftgithub}.

We present a generic framework for  simulation-based testing of ADS, and use this framework to perform experiments that answer RQ1 and RQ2.  The framework's test automation employs a basic random testing algorithm.   We choose random testing as the basis of our experiments because most ADS testing research relies on metaheuristic and fuzz testing algorithms that are fundamentally rooted in random testing methods~\citep{fuzzingbook2023:Coverage,metaheuristicsbook}. We assess test outputs using quantitative fitness functions that are defined based on system requirements.  These functions determine the extent to which a test satisfies or violates a given requirement. Fitness values are used to both guide search algorithms and generate Boolean verdicts, i.e., pass and fail results, for the requirements based on user-defined thresholds.

We define two distinct notions of flakiness for ADS testing: one based on Boolean verdicts and the other based on quantitative fitness values. The first notion, which  we refer to as \emph{hard flaky}, aligns with the definition of flaky tests in the literature:  A test is flaky if it passes and fails non-deterministically over multiple re-executions~\citep{survey}.  The second notion,  which we refer to as \emph{soft flaky}, identifies a test as flaky when there are variations in the values of a fitness function used for testing.

\textbf{Contributions.} We present the first study investigating the prevalence of flaky tests in ADS simulation-based testing and their impact on test results of ADS. Further, we study the effectiveness of machine learning classifiers in cost-effectively predicting flaky ADS tests and their ability to reduce the impact of flaky simulations on test results through a minimal number of test reruns. We address RQ1 and RQ2 (described earlier) using five distinct ADS test setups. Three of our test setups are adopted from the literature~\citep{samota,transfuser,haq2022manyobjective,sbftgithub}. We developed the two others to augment our empirical results. Our findings for RQ1 and RQ2 are summarized below:

\emph{RQ1)} Our results show that for our five test setups,  $4$\%-$68$\% of the generated tests exhibit notable variations in fitness values, indicating that all of our setups yield substantial soft flakiness. Further, the hard flaky rate for these setups ranges from $1$\% to $74$\%, with four out of five exhibiting a hard flaky rate exceeding $6$\% for at least one fitness function.  To assess the impact of flakiness on randomized ADS testing, we compare a random testing algorithm that captures the best fitness value from multiple candidate test reruns with a baseline random testing algorithm that executes each candidate test once. Our results show that the former substantially outperforms the latter, as it computes significantly better fitness values and identifies considerably more failures, ranging from $12$ to $888$.

%Given that these two metrics are generally used to compare the performance of different ADS testing methods~\citep{survey} our results show that flaky ADS tests may indeed alter our assessment of different randomized testing methods. 

\emph{RQ2)} To address the research question, we use three test setups, while excluding the Transfuser-based setup~\citep{transfuser} due to  its   prohibitive computational cost, and the competition-based setup~\citep{sbftgithub} as a result of its simple test input design and virtually negligible hard flaky rate. We build ML classifiers that can effectively identify flaky ADS tests  using only  a single test run,  achieving F1-scores of $85$\%, $82$\% and $96$\% for our three ADS test setups.  Using a non-ML baseline that detects flaky tests based on two or more test reruns, we show that our classifiers significantly outperform this baseline, achieving F1-score improvements of $31$\%, $21$\%, and $13$\%, respectively. 

In addition to answering the above two RQs, we present the following three lessons learned based on our findings: First, we confirm that the ADS test setup which is limited to checking the lane-keeping function~\citep{sbftgithub} shows the lowest rate of flaky tests. Second, based on our results, Pylot, which is a modular DNN-enabled ADS~\citep{gog2021pylot}, produces considerably lower flaky tests compared to Transfuser, which is an end-to-end ADS~\citep{transfuser}.  Third, the Carla simulator yields a lower flaky test  rate compared to the BeamNG simulator.

It is important to clarify that our results  should not be interpreted as criticisms of ADS simulators we consider in our study. CARLA and BeamNG are widely-used open-source simulators for ADS testing. They both have been used in several recent research on ADS testing, e.g.,~\citep{samota,zhong,haq2022manyobjective}. BeamNG has been used as a benchmark by the community~\citep{beamng}. As we discuss in Section~\ref{sec:conclusion}, many factors may contribute to the flakiness of simulators. Some of these factors, such as uncertainties in the physical models of simulators, are inherent to the physics-based design of both commercial and open-source simulators, and hence may be inevitable.\\
The main conclusion of our work is that the methods employed in the software engineering literature for assessing simulation-based ADS testing approaches may  lack robustness and may be unreliable  due to the flakiness of ADS simulators. Specifically, we rely on the number of individual failing test scenarios and the values of fitness functions as metrics to evaluate testing approaches. Our experiments, which are grounded in existing ADS test setups from the literature ~\citep{haq2022manyobjective,samota,sbftgithub}, demonstrate that the values of these metrics  can exhibit significant variability between two versions of baseline random testing: one that executes each test only once and another that repeats each test multiple times.  As we discuss in Section~\ref{sec:conclusion}, we either need to consider restricted ADS test setups where the impact of flakiness is minimized, or we need to develop  evaluation  metrics that are not sensitive to flakiness.

%Overall, our study reveals some limitations of ADS simulators and controllers and  helps identify conditions on test inputs and characteristics of ADS controllers that contribute to flakiness in this domain. We conclude the paper by reflecting on our observations and discussing the scope and implication of our study.

%To facilitate the replication of our study, we have made all the experimental  materials publicly available~\citep{github}.

\textbf{Organization.} Section~\ref{sec:example} motivates our work. Section~\ref{sec:background} describes our generic ADS testing framework. Section~\ref{sec:eval} presents our empirical study. Section~\ref{sec:relatedwork} compares with the related work. Section~\ref{sec:conclusion} outlines our observations and our discussions on the scope and implications of our study, and Section~\ref{sec:con} concludes the paper.

% !TEX root =  ../main.tex
\section{Examples of Flaky Simulations}
\label{sec:example}
\begin{figure}[t]
    \centering
    \includegraphics[width=\columnwidth]{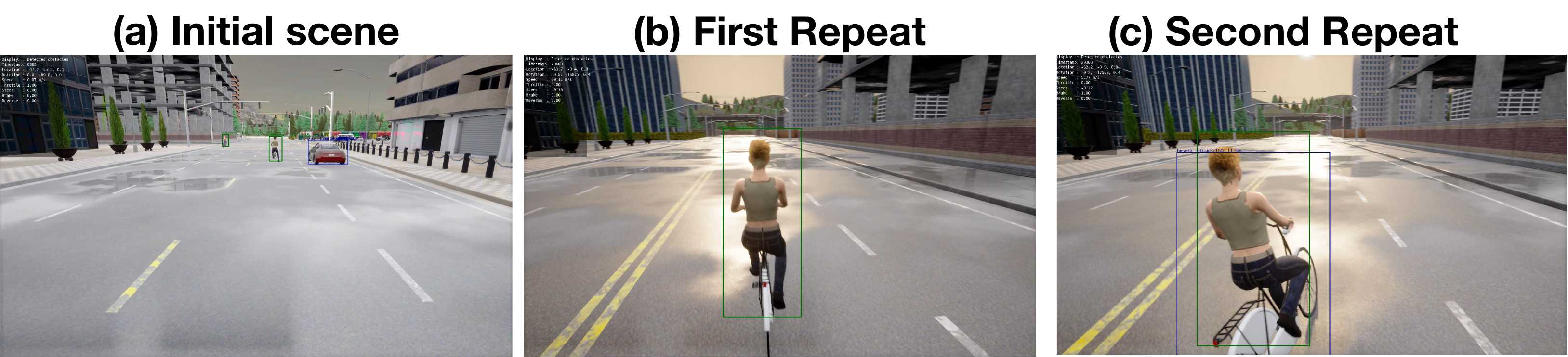}
    \caption{\protect \raggedright 
 Flakiness in ADS testing: Scene (a) is an initial scene, while  scenes (b) and (c) are taken from two separate  re-executions of the same test input starting from scene (a). Scenes (b) and (c) show the points of closest proximity between the ego car and the front bike. In scene (b), the accident is avoided, while  scene (c) shows an accident.}
    \label{fig:figflaky}
\end{figure}

 Figure~\ref{fig:figflaky} shows an example of flakiness in the CARLA simulator when executed with Pylot as the ADS. Figure~\ref{fig:figflaky}(a) shows the initial scene, and Figures~\ref{fig:figflaky}(b) and (c) show the moments of nearest distance between the ego car and the bicycle ahead during two distinct re-executions, both originating from scene~(a). These snapshots are captured by a camera mounted on the ego vehicle. In this example, Figure~\ref{fig:figflaky}(b) shows no accident between the ego car and the bike,  whereas Figure~\ref{fig:figflaky}(c) shows an accident between the ego car and  the bike.

% Flakiness in simulation-based ADS testing refers to the inconsistencies observed in test outputs when the same test input is re-executed multiple times. For example,

\begin{figure}[t]
    \centering
    \includegraphics[width=\columnwidth]{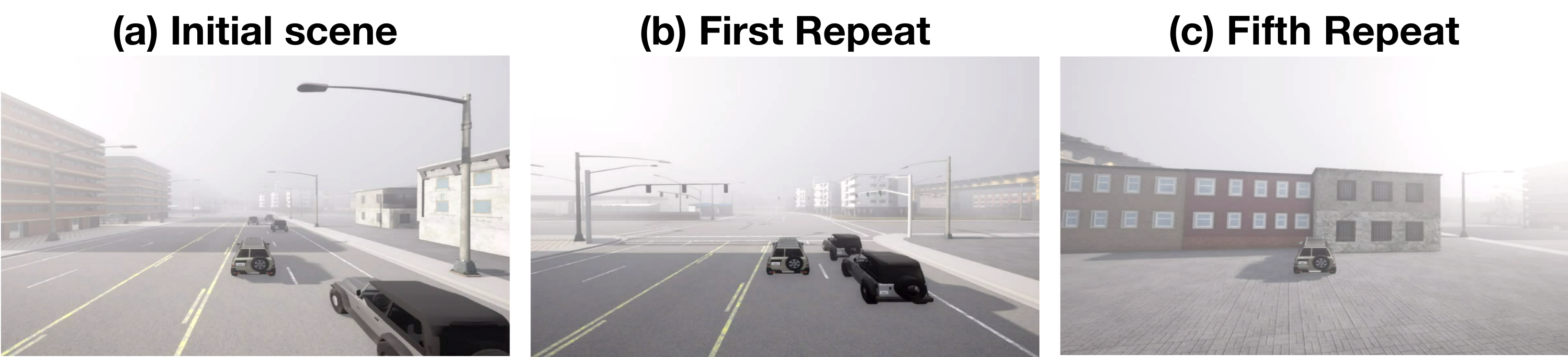}
    \caption{\protect \raggedright 
Flakiness in ADS testing. Similar to Figure~\ref{fig:figflaky}, scene~(a) is an initial scene, and  scenes (b) and (c) are from two re-executions starting from scene (a). The re-execution related to (b) shows no significant event, while in the re-execution represented by (c), the ego-car shows  unexplained behaviour and diverts from the road.}
    \label{fig:figflakyhard}
\end{figure}

The two re-executions shown in Figure~\ref{fig:figflaky} individually look like valid and realistic simulations. Sometimes, re-executions of the same test input represent rare or even impossible situations in the physical world. For example, Figure~\ref{fig:figflakyhard} shows another example of flakiness in ADS simulation-based testing obtained from CARLA executed with its PID-based ADS. While the re-execution represented by the scene in  Figure~\ref{fig:figflakyhard}(b) appears to be normal with no accidents or failures, the re-execution related to scene~(c) shows the ego car displaying abnormal and unexplained instability, veering off the road, entering a parking lot, and crashing into a building. Through our study for RQ1 and RQ2 described in Section~\ref{sec:intro}, we introduce an approach for evaluating and mitigating flakiness in ADS simulators. Our study uses an ADS simulation-based testing framework which is described in the next section.  

%This example also shows soft flakiness since there is considerable variations in the values of the fitness function measure the distance between the ego car and the bike between the two re-executions. 

\section{ADS Simulation-Based Testing}
\label{sec:background}
\begin{figure}[t]
    \centering
    \includegraphics[width=\columnwidth]{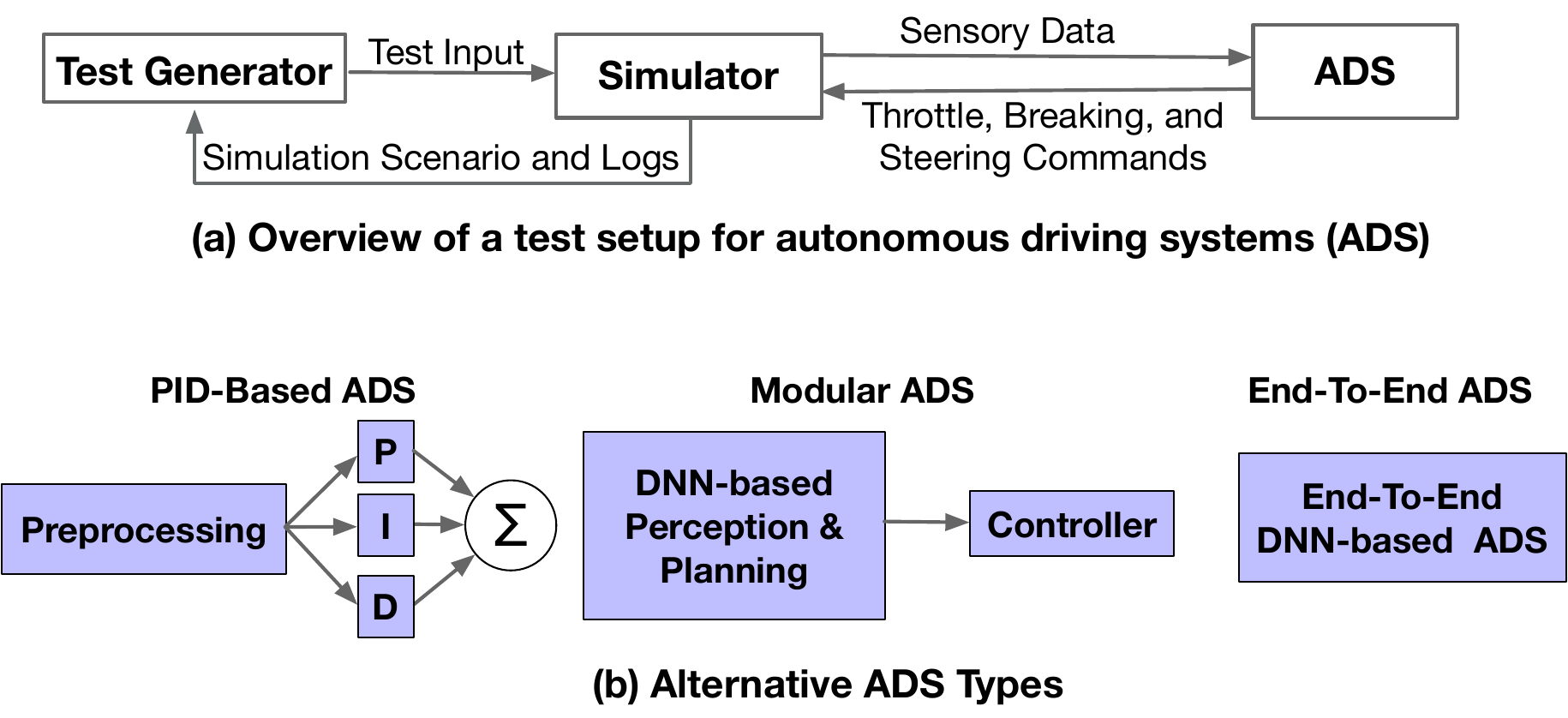}
    \caption{ADS simulation-based testing}
    \label{fig:sut}
\end{figure}

Figure~\ref{fig:sut}(a) shows an overview of ADS simulation-based testing that includes three elements: the test generator, the simulator, and the ADS which is the system under test (SUT). Below, we first provide background on each element. We then present a basic random testing algorithm for ADS. Finally, we define our two notions of flakiness in ADS testing.

%\begin{subfigure}{\columnwidth}
%        \centering
%        \includegraphics[width=\columnwidth]{figs/SUT.pdf}
%        \caption{Overview of a test setup for ADS testing}
%        \label{fig:sut-bigpicture}
%    \end{subfigure}
%    \begin{subfigure}{\columnwidth}
%        \centering
%        \includegraphics[width=\columnwidth]{figs/SUT-ADS.pdf}
%        \caption{Alternative ADS Architectures}
%        \label{fig:sut-ads}
%    \end{subfigure}

%we mean that some additional information is provided as to how objects behave during the simulation. In particular, the test input typically describes the initial and end points for the ego vehicle. 

\begin{figure}[t]
    \centering
    \includegraphics[width=\textwidth]{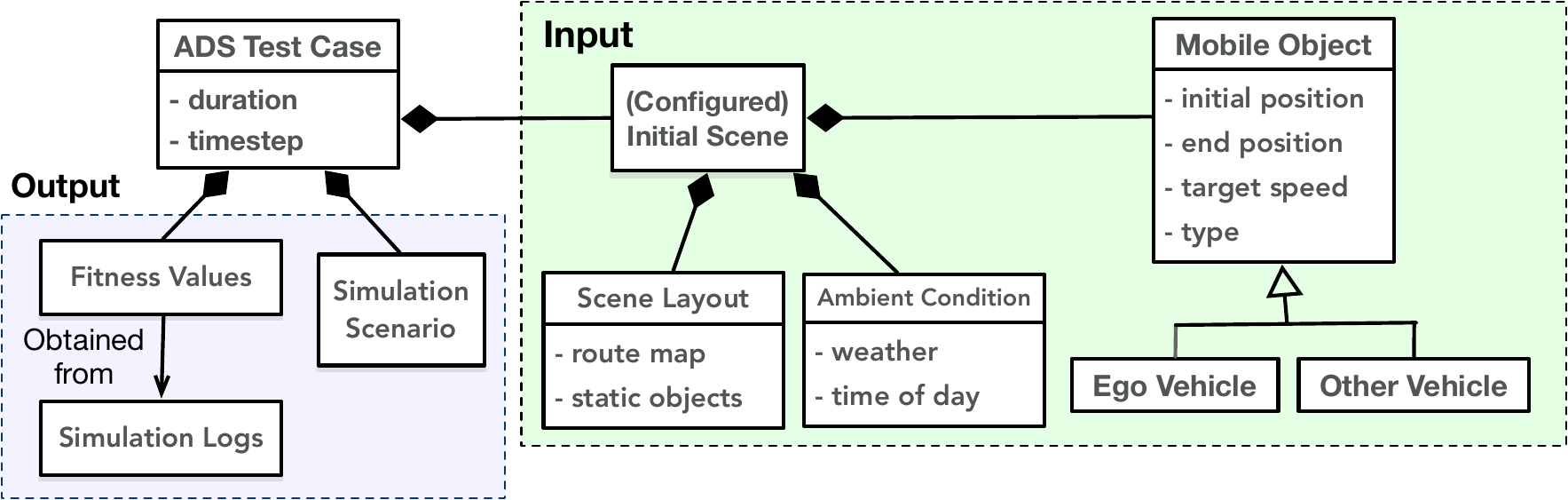}
    \caption{A conceptual model detailing test inputs and outputs used in our ADS test setups.}
    \label{fig:concept}
\end{figure}

\textbf{Test generator.} The test generator produces test inputs to be executed by the simulator and receives, as test outputs, simulation logs and a simulation scenario. The terms \emph{scene} and \emph{scenario} are often used in the ADS testing literature and are defined as follows: \emph{Scene} is a snapshot or frame in the simulation~\citep{zhong}, characterized by the properties of mobile objects, surrounding objects, roads and  ambient conditions. \emph{Scenario} is ``the temporal development between several scenes in a sequence of scenes''~\citep{definitionpaper}. Figure~\ref{fig:concept} shows a conceptual model detailing the test inputs and outputs that we designed for our ADS test setups. Specifically, a test input includes (1)~a \emph{configured} initial scene, (2)~simulation duration, and (3)~the time step duration, which is the time duration between each two consecutive scenes in a scenario. The configured initial scene includes the following information: (i)~Mobile objects that always include a single ego vehicle and optionally some non-ego vehicles. For each mobile object, we typically specify the initial and end points, the target speed and the vehicle type. (ii)~Surrounding objects such as buildings, traffic signs, parked cars, and pedestrians on the sidewalk. (iii)~Layout information such as route maps and road shapes. (iv)~Ambient conditions which include the weather condition and the time of the day. To be consistent with the ADS test setups adopted from the literature~\citep{samota, transfuser, haq2022manyobjective}, pedestrians are  static objects on the sidewalk. 
Due to the  presence of non-ego vehicles, we can still test  collision avoidance requirements using  our test setups.

% Remove Pedestrian. Explain how the information about the Vehicle and ego Vehicle is provided to the random tester. 

%Some of the properties in the initial scene remain constant during all the subsequent scenes in a simulation, and some properties change. We refer to the former as \emph{variable properties} and the latter as \emph{fixed properties}.  In ADS testing, the variable properties of a scene includes the location, velocity, acceleration, and orientation for the ego car and every mobile object, while other properties are typically fixed. Specifically, to test ADS,  we do not typically change the road condition, the weather or the ambient light situation within a single simulation scenario.  The set of properties (both fixed and variables) can be further decomposed based on the source that controls them. There are three sources that control scene properties: (i)~properties controlled by test inputs, (ii)~properties controlled by the ego vehicle (SUT), and (iii)~the properties controlled by simulator. Test inputs control a subset of the properties of the initial scene, but usually have no direct impact on the properties of the subsequent scenes within a simulation. The ego car controls the properties of the ego vehicle in all the scenes of a simulation scenario (except for the initial scene). Note that, the ego car operates in a closed-loop way, and at each scene, it acts upon the inputs it receives at the previous scene. The simulator often controls the properties of other moving objects. Depending on the simulator setup or its configuration, the properties of other moving objects may or may not change from one simulation to another. 

\textbf{Simulator.}  As Figure~\ref{fig:sut} shows, a simulator takes an initial configured scene as input and generates a scenario based on the specified  simulation time step and maximum duration.

\textbf{ADS.} The ADS receives sensory data from the simulator and generates the throttle, breaking and steering commands.  As Figure~\ref{fig:sut}(b) shows, we identify three different types of ADS based on their internal design: The first type primarily consists of a PID controller  which is combined with a preprocessing component responsible for perception and planning. This component may use non-DNN-based machine learning~\citep{carlatm, pidpaper}. The second type is a DNN-based modular design that integrates multiple DNNs into the perception and planning layer of ADS~\citep{gog2021pylot}. The DNN outputs are then passed to a controller that generates throttle, braking, and steering commands. The third type is a DNN-based end-to-end design that uses a single DNN for vehicle control. The DNN directly generates commands to be sent to the simulator~\citep{transfuser, udacity:challenge}. In our experiments, we use instances of each of these three types of ADS that have been previously used in the ADS testing literature~\citep{carlatm, gog2021pylot, transfuser,haq2022manyobjective,beamng,sbftgithub}.

\textbf{Fitness functions.} Fitness functions quantitatively estimate how close a test input is into violating the requirements of an ADS under test.  For example, collision avoidance can be measured by calculating the minimum distance between the ego and non-ego cars or static objects. We define binary  pass/fail verdicts to determine whether, or not, a test input violates a given requirement by comparing the fitness function value with a threshold. In our work, we adopt the thresholds from the prior studies that are often set at zero. For example, a test input violates the collision safety requirement if the minimum distance between the vehicles, or the ego car and an object is zero or near-zero.

\textbf{Random testing for ADS.} 
Most ADS testing research relies a search-based testing (SBT) or fuzz testing (FT) algorithms~\citep{fuzzingbook2023:Coverage,metaheuristicsbook,MatinnejadNB17}. SBT and FT  aim to generate  a limited and effective set of test cases using different meta-heuristics. Algorithm~\ref{alg:RS} shows a random testing algorithm which is the most basic form of SBT and FT. It randomly generates a test input $i$, and stores $i$'s fitness as the optimal fitness in $f_{\mathit{opt}}$ if it is better than the best fitness found so far. The algorithm continues until some stop condition is met. We assume  optimal tests are those that have the lowest fitness values. Due to flaky simulations, Algorithm~\ref{alg:RS} re-executes each test input $i$ for $n$ times by calling the   \emph{Fitness(i,n)} routine on line~7. Algorithm~\ref{alg:fitnessn} shows   \emph{Fitness(i,n)} that returns the most optimal fitness value (in our case, the lowest fitness value) among multiple re-executions of a given test input.  As discussed in Section~\ref{sec:intro}, we use Algorithm~\ref{alg:RS},  a basic random testing,  to assess the impact of flaky simulations on ADS testing. 

\begin{algorithm}[t]
%\scriptsize
\DontPrintSemicolon
\SetSideCommentRight
\SetNoFillComment
\textbf{Input: }$n \gets \text{Number of re-execution of each test input}$

\Begin{
$i \gets generateRandomTestInput()$

$f_{\mathit{opt}} \gets Fitness(i, n) $

\While{not (stop-condition)}{
$i \gets generateRandomTestInput()$

$f^{\prime} \gets Fitness(i, n) $

\If{$f^{\prime} < f_{\mathit{opt}}$} 
{
$f_{\mathit{opt}} \gets f^{\prime}$ 
}
}
\Return{$f_{\mathit{opt}}$}
}
\caption{Random Search}
\label{alg:RS}
\end{algorithm}

%a more critical test scenario. The fitness values, in addition to determining pass and fail or the degree severity, are also used to guide the test input sampling in the input space. Many explorative or exploitative search strategies are proposed in the literature that rely mainly on fitness values to determine which candidate test inputs should be considered in each search iteration.

\textbf{Flakiness definitions for ADS testing.} 
Using quantitative fitness values, flakiness can be measured in two ways: (1) soft flakiness, reflecting variations in fitness values across test re-executions, and (2) hard flakiness, indicating changes in pass/fail verdicts based on different test re-executions. Hard flakiness resembles flakiness in software code testing, while soft flakiness is an additional concept stemming from quantitative fitness functions. By definition, hard flakiness implies soft flakiness, but not vice versa.

%Show the random testing algorithm and argue that random testing is the most primitive search-based and fuzz testing algorithm. That's why we consider it.

%We show two variations of RS, with repeat and with discard corresponding to the two metrics used for assessing SBT and fuzz testing. 

\begin{definition}~\label{def:sf-hf} Let $i$ be a test input, and let $f_1 \ldots, f_n$ be the fitness values obtained from multiple executions of $i$. We define soft flakiness of $i$, denoted by $\mathit{SF}_i$, and hard flakiness of $i$, denoted by $\mathit{HF}_i$ as follows:

\scalebox{0.95}{$\begin{array}{l}
\label{eq:SF}
\mathit{SF}_i = \mathit{max}(\{f_1, \ldots, f_n\})-\mathit{min}(\{f_1, \ldots, f_n\})\\
\label{eq:HF}
\mathit{HF}_i = \exists{f, f'} \in \{f_1, \ldots, f_n\} : f > \mathit{thr} \wedge f' \leq \mathit{thr}
\end{array}$}

where the threshold \emph{thr} determines the pass/fail verdict. 
\end{definition}
Note that $\mathit{SF}_i$ is a quantitative measure, while $\mathit{HF}_i$ is Boolean. We use both notions of flakiness in our empirical evaluation presented in the next section.

%Simulators flakiness leads to variations both in the fitness values, i.e.,  soft flakiness, and in the number of reported failures, i.e., hard flakiness.  In order to assess the impact of ADS test flakiness on both the randomized heuristics utilized by SBT and FT algorithms and the metrics used for their assessment, we consider a basic random search algorithm shown in Algorithm~\ref{alg:RS}. 

\begin{algorithm}[t]
%\scriptsize
\DontPrintSemicolon
\SetSideCommentRight
\SetNoFillComment
\textbf{Input: }$i \gets \text{Test input}$

\textbf{Input: }$n \gets \text{Number of re-executions}$

\Begin{
$s \gets simulate(i)$

$f_1, f_{\mathit{opt}} \gets calculateFitness(s)$

\For{$j \in \{2, 3, \dots, n\} $}{
$s \gets simulate(i)$

$f_{j} \gets calculateFitness(s) $

\If{$f_{j} < f_{\mathit{opt}}$}
{
$f_{\mathit{opt}} \gets f_{j}$
}
}

\Return{$f_{\mathit{opt}}$}
}
\caption{Fitness(i, n)}
\label{alg:fitnessn}
\end{algorithm}

\section{Empirical Evaluation}
\label{sec:eval}

In this section, we study test flakiness in simulation-based ADS testing by answering the two research questions we motivated in Sections~\ref{sec:intro}, which are re-stated below:

\textbf{RQ1.} \emph{How do flaky ADS simulations impact automated testing that relies on randomized algorithms?} We study the impact of flaky ADS simulations on automated testing using three sub-research questions.  First,  we use the following sub-research question to determine the frequency of flaky tests: 

\begin{description}

\item[RQ1-1.] \emph{What is the frequency of flaky tests in ADS simulation-based testing?} We apply to five ADS test setups (i.e., Carla PID, Carla Pylot, Carla Transfuser, BeamNG AI, and BeamNG Competition) the random testing algorithm, Algorithm~\ref{alg:RS}. To report the soft and hard flakiness metrics (Definition~\ref{def:sf-hf}) for the randomly generated test,  each test is re-executed multiple times (i.e., in Algorithm~\ref{alg:RS}, we set $n>1$). 

\end{description}

A possible strategy is to discard test re-executions exhibiting extreme abnormalities. For example, Figure~\ref{fig:figflakyhard}(c) shows a test re-execution with extreme abnormality, while the test re-execution in Figure~\ref{fig:figflakyhard}(b) looks normal. Eliminating abnormal cases may mitigate flakiness if the majority of flaky ADS tests consist of occasional major abnormalities among predominantly normal and consistent re-executions. However, if most re-executions are normal  but still exhibit noticeable variations, such as the example in Figure~\ref{fig:figflaky}, this approach will not be effective. We present the following sub-research question to determine if most flaky tests are, in general,  categorized by consistent and normal re-executions with occasional extreme abnormalities, or if they exhibit variations but with few extreme abnormalities:

\begin{description}

\item[RQ1-2.] \emph{Do individual re-executions of flaky ADS test inputs represent normal scenarios?}  We examine re-executions of flaky tests to determine whether they represent  extreme abnormalities, or whether, despite generating different fitness values, they still represent normal driving scenarios.
\end{description}

In addition to evaluating the prevalence of flaky tests (RQ1-1) and examining the normality of the observed variations over test reruns (RQ1-2), we evaluate the scale of the impact of flaky tests on ADS testing through the following question:

\begin{description}

\item[RQ1-3.] \emph{Can the variations caused by flaky tests substantially impact the results of ADS testing algorithms?}  We compare the performance of random testing, a baseline ADS testing algorithm,   for two different configurations: (1)~each test input candidate is executed once (i.e., Algorithm~\ref{alg:RS} with $n=1$), and (2)~each test input candidate is executed multiple times (i.e., Algorithm~\ref{alg:RS} with $n>1$).  For a deterministic test setup, rerunning the same candidate tests  multiple times should have no impact.  While the presence of flaky tests means that rerunning the same test is likely to impact the results of ADS testing algorithms, ideally, the magnitude and significance of this impact should be minimal.
We evaluate the significance of observed variations using effect sizes and statistical tests for the two widely-used metrics in the SBT and FT literature~\citep{raja2018, samota, fuzzingbook2023:Coverage}: the number of detected failures and the optimality of the fitness values calculated by each algorithm.

%Further, we define a quantitative function for each observed variation such that the value of the function determines if the  variation is present in a given scenario, and further, the function can estimate the extent of the variation.  We then compare the variations and see if there is any shared observed variation and how many shared observed variations exist. 

\end{description}

\textbf{RQ2.} \emph{Can machine learning (ML) effectively identify flaky ADS tests while decreasing the required number of test reruns?} Since both soft and hard flakiness may impact ADS testing, and since soft flaky is a weaker notion, we focus on using ML to predict soft flakiness for ADS test inputs. To do so, we use the set of input variables and fitness values as features for learning. These features are typically available for any ADS test setup regardless of the specific simulator or the ADS controller used. We experiment with different subsets of input variables to identify the most relevant and informative features for predicting flakiness. In addition, we consider two alternative feature designs to capture fitness functions. The first design uses individual fitness values obtained from single test runs, while the second design uses differences of fitness values obtained from multiple re-executions of the same tests. The first design requires less effort as it only needs one execution of each test input, while the second design needs multiple executions. We refer to an ML model built based on the first design as single-test-execution classifier (\Ofit) and that built based on the second design as multi-test-execution classifier (\Odif). We assess both types of classifiers by studying the trade-off between their prediction accuracy and the number of (re-)executions needed to produce their input features. We also compare their performance with a baseline that does not rely on machine learning.

Predicting flakiness in ADS testing  may fulfill two different goals. The first goal concerns with detecting flaky test inputs, while the second aims to improve ADS testing by identifying failure scenarios with the most optimal (lowest) fitness values. We examine these goals through the following sub-questions:

\begin{description}
\item[RQ2-1] \emph{How accurate and cost-effective are machine learning classifiers in predicting flaky ADS tests?} We evaluate the precision, recall and F-1 score of alternative  classification techniques in predicting flaky ADS test inputs. We consider the two feature designs (i.e, \Ofit and \Odif) to build classifiers. Further, we compare the classifiers with a non-ML baseline.

\item[RQ2-2] \emph{Can machine learning classifiers improve the performance of ADS testing by requiring a limited number of test reruns?} For this question, we modify Algorithm~\ref{alg:fitnessn} to develop Algorithm~\ref{alg:smartfitnessn}, which uses predictions of ML classifiers (i.e., \Ofit and \Odif) to minimize the number of required reruns for a candidate test. In other words, Algorithm~\ref{alg:smartfitnessn}, instead of re-running each test an equal number of times, reruns a test only until we can infer it is flaky. Following the first fitness calculation,  \Ofit predicts, in line 6 of Algorithm~\ref{alg:smartfitnessn}, whether test $i$ is flaky or not, based on the fitness value obtained from the single simulation in line 4. The algorithm proceeds with the while-loop in lines~8-15 only if \Ofit labels test $i$ as flaky. In each iteration, the while-loop re-executes test $i$, computes a new fitness value, and calculates the maximum difference  (i.e., $\delta$) among fitness values obtained for test $i$ so far. It then uses  \Odif to predict flakiness based on $\delta$. Note that inside the loop, given the availability of multiple test executions and the maximum fitness difference, we use  \Odif instead of \Ofit. The while-loop runs up to  $n$ iterations or 
until \Odif labels test $i$ as non-flaky. The algorithm concludes by returning the optimal fitness value ($f_{\mathit{opt}}$).

While Algorithm~\ref{alg:fitnessn} runs each candidate test $n$ times,  Algorithm~\ref{alg:smartfitnessn} likely runs several candidate tests only once or fewer than  $n$ times. Our goal is to experiment with different ML classifiers to determine if Algorithm~\ref{alg:smartfitnessn} can obtain optimal fitness values that are close to those obtained by  Algorithm~\ref{alg:fitnessn} while requiring far fewer test reruns. 

\end{description}
%Note that using $\delta$ as input for \Odif ensures the model interface and architecture remains independent of the number of re-executions. In addition, the training sets for  \Odif are also independent of the of the number of re-executions. For example, a training set with $n$ reruns for each test leads to  $\binom{n}{2}$ data points with potential duplicates and assuming that subsequent simulations are independent. it helps the model generalize better with less amount of data available.  

%However, there is a slight difference between soft flakiness and hard flakiness computation in \emph{Fitness(i,n)} and \emph{fastFitness(i,n)} routines. In the \emph{Fitness(i,n)} routine, fitnesses from execution $1$ to $n$ are used, while in the \emph{fastFitness(i,n)}, we don't necessarily re-execute n times. Indeed, the main idea of the \emph{fastFitness(i,n)} is to re-execute less than n times if we can. Assuming that for test input $i$, the number of re-executions in the \emph{fastFitness(i,n)} routine is $n_i$, the new definitions of soft flakiness and hard flakiness metrics would be just like equtions \ref{eq:SF} and \ref{eq:HF}, except that $n$ would be replaced by $n_i$ in them.

\begin{algorithm}[t]
%\scriptsize
\DontPrintSemicolon
\SetSideCommentRight
\SetNoFillComment
\textbf{Input: }$i \gets$ \text{Test input}

\textbf{Input: }$n \gets$ \text{Maximum Number of re-executions}

\Begin{

$s \gets simulate(i)$

$f_1, f_{\mathit{opt}} \gets calculateFitness(s)$

$\mathit{flaky} = \Ofit(i, f_1)$  // {\footnotesize\textcolor{javagreen}{single-test-execution classifier}}

$j \gets 1$

% \For{$j \in \{2, 3, \dots, n\} $}{
\While{(flaky $\wedge (j < n)$)}{

$j \gets j+1$
        
$s \gets simulate(i)$

$f_{j} \gets calculateFitness(s)$

\If{$f_{j} < f_{\mathit{opt}}$}
{
$f_{\mathit{opt}} \gets f_{j}$
}
$\delta = \max(\{f_1, \ldots, f_j\})-\min(\{f_1, \ldots, f_j\})$

$\mathit{flaky} = \Odif(i, \delta)$ // {\footnotesize\textcolor{javagreen}{multi-test-execution classifier}}

}
\Return{$f_{\mathit{opt}}$};
}
\caption{fastFitness(i, n)}
\label{alg:smartfitnessn}
\end{algorithm}

%The \emph{smartFitness(i,n)} algorithm assumes that an oracle $\mathcal{O}$ is available that predicts the flakiness for a test input $i$. The prediction can be done based on the test input alone, or based on the test input and the differences between the fitness values obtained from one or multiple of re-executions of the test input. \textcolor{blue}{why is this a good idea?}. \textcolor{blue}{We want the oracle to use the data that is available for ADS setting.}

\subsection{ADS test setups} 

Our study uses the CARLA~\citep{carlapaper} and BeamNG~\citep{beamng} simulators together with four different ADS: (1)~CARLA's PID-based traffic manager~\citep{carlatm}; (2)~Pylot, a DNN-enabled modular ADS~\citep{gog2021pylot}; (3)~Transfuser, a DNN-enabled end-to-end ADS~\citep{transfuser}; and (4)~BeamNG's AI-driven default ADS~\citep{beamng}. We develop five different instances of the ADS test setup of Figure~\ref{fig:sut} by combining  these simulators and ADS.  Three of these setups use CARLA as the simulator, with traffic manager, Pylot, and Transfuser as the respective ADS. In the remainder of this paper, we refer to these three as \textsc{PID}, \textsc{Pylot} and \textsc{Tran}, respectively. For the test inputs and fitness functions of \textsc{Pylot} and \textsc{Tran}, we rely on the test generators provided in the literature~\citep{samota,transfuser,haq2022manyobjective}. We  ensure the consistency of  the test generators for these two setups. The details as to how we adopted the test inputs and fitness functions for these two setups are  available in our repository~\citep{github}. For \textsc{PID}, we developed, from scratch, a test generator that produces test inputs and computes fitness functions compatible with those used for \textsc{Pylot} and \textsc{Tran}. For the other two test setups, we use BeamNG as the simulator together with its AI-based ADS, but we use two different test generators: one that aligns with the test generators for \textsc{PID}, \textsc{Pylot} and \textsc{Tran}, and another based on the setup for the Cyber-Physical Systems Testing Tool Competition track of the SBFT workshop~\citep{sbftgithub}. We refer to these two setups as \textsc{BeamNG} and \textsc{Comp}, respectively. The characteristics of our ADS test setups are summarized in Table~\ref{tab:expinfo}.
Below, we briefly describe the test inputs and fitness functions used in our five test setups.

\begin{table}[t]
\begin{center}
\caption{Characteristics of the ADS test setups used in our experiments.}
\label{tab:expinfo}
\scalebox{.93}{
\begin{tabular}{|p{1.1cm}|p{1.4cm}|p{1.5cm}|p{2cm}|p{1.5cm}|p{2cm}|}
\hline
\textsc{Setup ID} & \textsc{Simulator} & \textsc{ADS} & \textsc{Test Inputs} & \textsc{Fitness Functions}  & \textsc{source} \\ \hline

\textsc{PID} & CARLA & CARLA PID & Figure~\ref{fig:concept} & $F_1$ \ldots $F_4$ & Our replication package~\citep{github}\\ \hline

\textsc{Pylot} & CARLA & Pylot& Figure~\ref{fig:concept} & $F_1$ \ldots $F_4$ & \citep{gog2021pylot}\\ \hline

\textsc{Tran}& CARLA & Transfuser& Figure~\ref{fig:concept}& $F_1$ \ldots $F_4$  & \citep{transfuser, samota}  \\ \hline

\textsc{BeamNG} & BeamNG& BeamNG AI & Figure~\ref{fig:concept}&$F_1$ \ldots $F_4$  & Our replication package~\citep{github}\\ \hline

\textsc{Comp} & BeamNG &BeamNG AI & A single-lane road & $F_1$ & \citep{sbftgithub} \\ \hline

\end{tabular}
}
\end{center}
\vspace*{-.5cm}
\end{table}

\textbf{Test inputs.} The test inputs for \textsc{PID}, \textsc{Pylot}, \textsc{Tran} and \textsc{BeamNG} conform to the conceptual model in Figure~\ref{fig:concept}.  %Specifically,  \textsc{Pylot} and \textsc{PID} use a fixed route map for all the test inputs, while \textsc{Tran} uses different route maps. In addition our test inputs may include upto ten non-ego vehicles with specified types, speed, and  initial and  end positions. 
% \textcolor{blue}{Out of these four, mention that two setups are reused from the literature, and two setups are developed by us to be consistent with them}.
%Out of these four test setups, \textsc{Pylot} and \textsc{Tran} are used in the same configuration as the literature \citep{gog2021pylot, transfuser}, while \textsc{PID} and \textsc{BeamNG} are developed from scratch by the authors.
For \textsc{Comp},  we follow the competition website's test input design~\citep{sbftgithub}, which only includes information about the route map. \textsc{Comp} test inputs do not include any non-ego vehicles, static objects or any information about the weather or the time of day. 
The \textsc{Pylot}, \textsc{Tran} and \textsc{Comp} repositories~\citep{gog2021pylot, transfuser, samota, sbftgithub} already include a random testing baseline implemented. This random testing samples test inputs within their specified ranges assuming that each input variable has a uniform distribution. We used these already implemented random baselines for our experiments, and implemented similar random testing algorithms for  \textsc{PID} and \textsc{BeamNG}, i.e., the test setups implemented in this paper. The implementation of the random testing algorithms is available  in our replication package [2].

\textbf{Fitness functions.} For \textsc{PID}, \textsc{Pylot}, \textsc{Tran} and \textsc{BeamNG}, we evaluate the test outputs against four ADS requirements: 

\begin{description}
\item[R1:] \emph{The ego car should remain within its lane, only deviating when intentionally changing lanes}.
\item[R2:] \emph{The ego car should always maintain a safety distance from other vehicles}. 
\item[R3:] \emph{The ego car should always maintain a safety distance from the sidewalk and static objects}. \item[R4:] \emph{The ego car should reach the specified destination  within the maximum simulation time duration}. 
\end{description}

We define four fitness functions, referred to as $F_1$, $F_2$, $F_3$ and $F_4$, respectively, to evaluate these four requirements:

\begin{description}
\item[$F_1$] measures the number of lane invasions not followed by a lane change. 

\item[$F_2$] measures the minimum distance between the ego and non-ego cars; 
\item[$F_3$] measures the minimum distance between the ego car and static objects or sidewalk; and 
\item[$F_4$] measures the distance to the destination position.
\end{description}

For \textsc{Pylot}, \textsc{Tran} and \textsc{Comp}, the implementations of the above four fitness functions are respectively taken from the repositories provided by the sources of these  ADS test setups~\citep{samota,haq2022manyobjective,sbftgithub}. 
The Comp setup~\citep{sbftgithub} uses a one-lane road. For this setup,  a lane invasion is computed based on  the out-of-bound (OOB) distance  that is measured as  the average of the differences between the lane’s width and the distance from the ego car to the lane’s center across all time steps of the simulation scenario. For \textsc{Pylot}~\citep{samota} and \textsc{Tran}~\citep{haq2022manyobjective}, $F_1$ is calculated similarly to OOB. However, a distinction is made considering that the map for \textsc{Pylot} and \textsc{Tran} is a two-lane road intersection, and the ego car must make a lane change to reach its destination. In this context, one lane invasion followed by a lane change is deemed intentional and is not included in the computation of $F_1$. For \textsc{BeamNG} and \textsc{PID} that are implemented by the authors, we have adopted this latter implementation for $F_1$ since the map of \textsc{BeamNG} and \textsc{PID} is similar to that of \textsc{Pylot} and \textsc{Tran}. 
We have also adopted the implementations of  $F_2$, $F_3$ and $F_4$ from the \textsc{Pylot}~\citep{samota} setup and 
verified their consistency with those in the Tran setup~\citep{haq2022manyobjective}.  Finally, for each fitness function, we used the thresholds provided by the source repositories for \textsc{Pylot}, \textsc{Tran} and \textsc{Comp}~\citep{samota, sbftgithub, transfuser}. The implementation of the four fitness functions and the threshold used for each fitness function are detailed in our replication package~\citep{github}. 
%to differentiate between pass and fail. These thresholds were determined independently of test results, relying solely on the function's underlying semantics and prior ADS testing studies~\citep{samota, sbftgithub, transfuser}.   

%\textbf{Transfuser.} \textcolor{blue}{What is the input scene and what are the fitness functions here?}

%\textbf{-} Deviations from the centre of the lane. 

%\textbf{-} Avoiding collision with other vehicles. 

%\textbf{-} Avoiding collision with pedestrians. 

%\textbf{-} Avoiding collision with static obstacles. 

%\textbf{-} Respecting traffic rules.

%\textbf{-} Distance travelled towards the destination.

%These elements include: (1)~the states of traffic lights,  (2)~Traffic manager, (3)~Synchronicity, and (4)~non-ego vehicles. Below, we discuss in what ways each element may introduce randomness, and how we have ensured to control them. 

%\subsection{Simulation configuration}  

%We ensure to determine each of the properties of non-ego vehicles at the beginning of each simulation.

%Despite all our efforts to keep parameters in control, our results suggest that there may be additional factors causing the simulator to behave in a flaky manner.

%\input{tex/tab.tex}

\subsection{RQ1-1 Results} 
This research question aims to identify the frequency of flaky tests in ADS simulation-based testing.
For this research question, we apply random testing (Algorithm~\ref{alg:RS}) to each of our five test setups. We set the number of test reruns, i.e., parameter $n$ of Algorithm~\ref{alg:RS}, to ten based on our preliminary experiments that were aimed at revealing flakiness in our different test setups. We generate $1000$ random tests for each of \textsc{PID}, \textsc{Pylot}, \textsc{BeamNG}, and \textsc{Comp}, and $100$ random tests for \textsc{Tran}. We ensure the generation of diverse and unique test inputs as per the definitions of test input diversity in ADS testing~\citep{raja2018,zhong}.  Each execution of \textsc{PID}, \textsc{Pylot}, \textsc{Tran}, \textsc{BeamNG} and \textsc{Comp} on average takes $1.5$min, $5.6$min, $12$min, $2$min, and $1$min, respectively. In total, we performed $10,000$ simulations for each of \textsc{PID}, \textsc{Pylot}, \textsc{BeamNG}, and \textsc{Comp}, and $1,000$ simulations for \textsc{Tran}.
% \linelabel{A_R3C9_l1}
%They all took $10$ days each, except for \textsc{Pylot} which took $40$ days. For \textsc{Tran}, $500$ simulations were done in about $20$ days. 
All experiments were conducted on a machine with a 2.5 GHz Intel Core i9 CPU and 64~GB of DDR4 memory.

%\emph{Experiment setup.} For the CARLA+Pylot platform, we executed a $50$ iteration random test algorithm for 20 times. We then executed a $50$ iteration random test such that, at each iteration, we re-executed the simulation for the selected test input ten times and picked the worst fitness value for that test input. In total, we performed $11000$ simulations for CARLA+Pylot platform. This took over a month. For CARLA+Transfuser, we used 50 randomly selected test inputs. Among these 36 were provided on the online github of Transfuser. We generated addition 14 scenarios randomly. We executed the 50 scenarios and similar to the CARLA+Pylot platform, we repeated the execution of each scenario ten times and recorded the fitness variations as well as the worst fitness.  The experiments for CARLA+Transfuser took about about 20 days to complete for the 50 scenarios. For the CARLA platform, we executed a $50$ iteration random test algorithm for $20$ times. We then executed a $50$ iteration random test such that, at each iteration, we re-executed the simulation for the selected test input ten times and picked the worst fitness value for that test input. In total, we performed $10000$ simulations for CARLA platform that took two weeks.

\textbf{Metrics.} We calculate the soft flaky $\mathit{SF}$  and hard flaky $\mathit{HF}$  measures (see Definition~\ref{def:sf-hf}) 
for every fitness $F_1$, $F_2$, $F_3$ and $F_4$ of each setup individually. Recall that the \textsc{Comp} setup has only one fitness function ($F_1$).

\textbf{Results.} Table~\ref{tab:tabrq1} presents the soft flakiness results for each fitness function of each setup.  Recall from Definition~\ref{def:sf-hf} that soft flakiness captures the variations in the fitness function values and is denoted by $\mathit{SF}_j$ for a test input $j$.  Further, we denote by $\mathit{MaxSF}_{s,F}$ the largest soft flaky value  among all the test inputs generated for setup $s$ and fitness function $F$, and by $R_{s, F}$ the value range of  the fitness function $F$ for setup $s$. The third and fourth columns from the left of Table~\ref{tab:tabrq1}, respectively, show the fitness range ($R$) and the max soft flaky value ($\mathit{MaxSF}$) for each fitness function of each setup. Among the 17 rows of Table~\ref{tab:tabrq1}, in twelve rows, the maximum soft flaky is equal to the fitness range; in three rows, $\mathit{MaxSF}$ is at least $93\%$ of the  fitness range; for  one case, $\mathit{MaxSF}$ is at least $80\%$ of the fitness range; and only in one case (i.e, $F_3$ of \textsc{Pylot}), $\mathit{MaxSF}$ is only $20$\% of the maximum fitness range. These results show that maximum soft flakiness can be as high the fitness function range in several cases. 

For each setup $s$ and each fitness function $F$, we compute the ratio of soft flakiness for test input $j$ as $\mathit{SF}_j/\mathit{MaxSF}_{s,F}$. The fifth to ninth columns from the left of Table~\ref{tab:tabrq1} show, for each fitness function and each setup, the number of test inputs with soft flaky ratios that fall into different intervals, ranging from $[0 - 1\%]$ up to $(40\% - 100\%]$.  For example, out of $1000$ test inputs generated for the fitness $F_1$ of \textsc{Pylot}, $443$ have soft flaky ratios within the 0-1\% range.
But, for the same fitness function and the same setup, the soft flaky ratio for $230$ test inputs is more than $5$\%. Considering the soft flaky ratio of more than $5$\% as non-negligible, at least $50$\% of the tests  in the six rows highlighted blue show non-negligible soft flakiness, and at least $10$\% of the tests  in the five rows highlighted green show non-negligible soft flakiness. %In the remaining six rows, at least $4$\% of tests show non-negligible soft flakiness. 

%Note that the data in Table~\ref{tab:tabrq1} is based on $1000$ test inputs for Pylot and PID, and $50$ test inputs for Transfuser. 

\begin{table}[t]
\centering
\caption{Soft flakiness (SF) results for each fitness function of our five test setups. The third and fourth columns from the left, respectively, show the fitness function range ($R$) and the maximum SF value ($\mathit{MaxSF}$). The fifth to ninth columns from the left show intervals for soft flaky (SF) ratios with $0\%$ meaning no soft flaky and $100\%$ being the maximum SF value (i.e., $\mathit{MaxSF}$). For example, the sixth column from the left indicates the number of test inputs with a soft flaky ratio ($SF/\mathit{MaxSF}$) within the $\mathbf{(1\%-5\%]}$ interval. Considering the soft flaky ratio of more than 5\% as non-negligible, rows with a minimum of 50\% and 10\% of tests exhibiting non-negligible soft flakiness are highlighted in blue and green, respectively.}
\label{tab:tabrq1}
\scalebox{0.78}{ \begin{tabular}{|p{1.1cm}|p{0.3cm}|p{0.5cm}|p{.8cm}|p{1.5cm}|p{1.5cm}|p{1.8cm}|p{1.8cm}|p{1.9cm}|} 
\hline
{\footnotesize\textbf{Setup}} & {\footnotesize\textbf{F.}}   & $R$                          & {\small$\mathit{MaxSF}$}     & {\footnotesize\textbf{[0\%--1\%]}} & {\footnotesize\textbf{(1\%--5\%]}} & {\footnotesize\textbf{ (5\%--10\%]}} & {\footnotesize\textbf{(10\%--40\%]}} & {\footnotesize\textbf{(40\%--100\%]}} \\ \hline
                              & \cellcolor[HTML]{FFFFFF}$F_1$     & \cellcolor[HTML]{FFFFFF}2    & \cellcolor[HTML]{FFFFFF}1.99 & \cellcolor[HTML]{FFFFFF}79\% (783) & \cellcolor[HTML]{FFFFFF}14\% (140) & 2.4\% (24)                           & 5\% (49)                             & 0.4\% (4)                            \\ \cline{2-9} 
                              & \cellcolor[HTML]{FFFFFF}$F_2$     & \cellcolor[HTML]{FFFFFF}2    & \cellcolor[HTML]{FFFFFF}1.79 & \cellcolor[HTML]{FFFFFF}6\% (58)   & \cellcolor[HTML]{FFFFFF}30\% (297) & \cellcolor[HTML]{38FFF8}16\% (157)   & \cellcolor[HTML]{38FFF8}26\% (260)   & \cellcolor[HTML]{38FFF8}23\% (228)   \\ \cline{2-9} 
                              & \cellcolor[HTML]{FFFFFF}$F_3$     & \cellcolor[HTML]{FFFFFF}2    & \cellcolor[HTML]{FFFFFF}2    & \cellcolor[HTML]{FFFFFF}77\% (765) & \cellcolor[HTML]{FFFFFF}4\% (37)   & \cellcolor[HTML]{32CB00}2\% (22)     & \cellcolor[HTML]{32CB00}15\% (151)   & \cellcolor[HTML]{32CB00}3\% (25)     \\ \cline{2-9} 
\multirow{-4}{*}{\textsc{PID}}         & \cellcolor[HTML]{FFFFFF}$F_4$     & \cellcolor[HTML]{FFFFFF}2    & \cellcolor[HTML]{FFFFFF}2    & \cellcolor[HTML]{FFFFFF}87\% (869) & \cellcolor[HTML]{FFFFFF}9\% (94)   & 2\% (16)                             & 2\% (15)                             & 1\% (6)                              \\ \hline
\textsc{Pylot}                         & \cellcolor[HTML]{FFFFFF}$F_1$     & \cellcolor[HTML]{FFFFFF}1.59 & \cellcolor[HTML]{FFFFFF}1.26 & \cellcolor[HTML]{FFFFFF}44\% (433) & \cellcolor[HTML]{FFFFFF}34\% (337) & \cellcolor[HTML]{32CB00}12\% (122)   & \cellcolor[HTML]{32CB00}8\% (82)     & \cellcolor[HTML]{32CB00}3\% (26)     \\ \cline{2-9} 
                              & \cellcolor[HTML]{FFFFFF}$F_2$     & \cellcolor[HTML]{FFFFFF}2.98 & \cellcolor[HTML]{FFFFFF}2.75 & \cellcolor[HTML]{FFFFFF}90\% (898) & \cellcolor[HTML]{FFFFFF}4\% (36)   & 1\% (10)                             & 4\% (37)                             & 2\% (19)                             \\ \cline{2-9} 
                              & \cellcolor[HTML]{FFFFFF}$F_3$     & \cellcolor[HTML]{FFFFFF}0.90 & \cellcolor[HTML]{FFFFFF}0.18 & \cellcolor[HTML]{FFFFFF}96\% (963) & \cellcolor[HTML]{FFFFFF}0\% (0)    & 1\% (7)                              & 1\% (10)                             & 2\% (20)                             \\ \cline{2-9} 
                              & \cellcolor[HTML]{FFFFFF}$F_4$     & \cellcolor[HTML]{FFFFFF}0.96 & \cellcolor[HTML]{FFFFFF}0.96 & \cellcolor[HTML]{FFFFFF}90\% (899) & \cellcolor[HTML]{FFFFFF}1\% (12)   & 1\% (6)                              & 4\% (41)                             & 4\% (42)                             \\ \hline
\textsc{Tran}                          & \cellcolor[HTML]{FFFFFF}$F_1$     & \cellcolor[HTML]{FFFFFF}1.11    & \cellcolor[HTML]{FFFFFF}1.11    & \cellcolor[HTML]{FFFFFF}0\% (0)  & \cellcolor[HTML]{FFFFFF}6\% (6)    & \cellcolor[HTML]{38FFF8} 21\% (21)                              & \cellcolor[HTML]{38FFF8} 43\% (43)                              & \cellcolor[HTML]{38FFF8} 30\% (30)                             \\ \cline{2-9} 
                              & \cellcolor[HTML]{FFFFFF}$F_2$     & \cellcolor[HTML]{FFFFFF}142   & \cellcolor[HTML]{FFFFFF}142   & \cellcolor[HTML]{FFFFFF}87\% (87)   & \cellcolor[HTML]{FFFFFF}3\% (3)  & \cellcolor[HTML]{32CB00}1\% (1)    & \cellcolor[HTML]{32CB00}7\% (7)    & \cellcolor[HTML]{32CB00}2\% (2)      \\ \cline{2-9} 
                              & \cellcolor[HTML]{FFFFFF}$F_3$     & \cellcolor[HTML]{FFFFFF}0.3    & \cellcolor[HTML]{FFFFFF}0.3    & \cellcolor[HTML]{FFFFFF}46\% (46)  & \cellcolor[HTML]{FFFFFF}50\% (50)    & \cellcolor[HTML]{FFFFFF}0\% (0)      & \cellcolor[HTML]{FFFFFF}0\% (0)      & \cellcolor[HTML]{FFFFFF}4\% (4)     \\ \cline{2-9} 
                              & \cellcolor[HTML]{FFFFFF}$F_4$     & \cellcolor[HTML]{FFFFFF}0.08 & \cellcolor[HTML]{FFFFFF}0.08 & \cellcolor[HTML]{FFFFFF}0\% (0)  & \cellcolor[HTML]{FFFFFF}2\% (2)    & \cellcolor[HTML]{38FFF8}20\% (20)      & \cellcolor[HTML]{38FFF8}66\% (66)     & \cellcolor[HTML]{38FFF8}12\% (12)    \\ \hline
                              & \cellcolor[HTML]{FFFFFF}\emph{F1} & \cellcolor[HTML]{FFFFFF}2    & \cellcolor[HTML]{FFFFFF}2    & \cellcolor[HTML]{FFFFFF}33\% (333) & \cellcolor[HTML]{FFFFFF}0\% (0)    & \cellcolor[HTML]{38FFF8}0\% (0)      & \cellcolor[HTML]{38FFF8}0\% (0)      & \cellcolor[HTML]{38FFF8}67\% (667)   \\ \cline{2-9} 
\multirow{-2}{*}{\textsc{BeamNG}}      & \cellcolor[HTML]{FFFFFF}$F2$ & \cellcolor[HTML]{FFFFFF}2    & \cellcolor[HTML]{FFFFFF}2    & \cellcolor[HTML]{FFFFFF}34\% (337) & \cellcolor[HTML]{FFFFFF}20\% (201) & \cellcolor[HTML]{32CB00}18\% (177)   & \cellcolor[HTML]{32CB00}28\% (276)   & \cellcolor[HTML]{32CB00}1\% (9)      \\ \cline{2-9} 
                              & \cellcolor[HTML]{FFFFFF}$F3$ & \cellcolor[HTML]{FFFFFF}2    & \cellcolor[HTML]{FFFFFF}2    & \cellcolor[HTML]{FFFFFF}24\% (240) & \cellcolor[HTML]{FFFFFF}8\% (77)   & \cellcolor[HTML]{38FFF8}2\% (24)     & \cellcolor[HTML]{38FFF8}34\% (335)   & \cellcolor[HTML]{38FFF8}32\% (324)   \\ \cline{2-9} 
                              & \cellcolor[HTML]{FFFFFF}$F4$ & \cellcolor[HTML]{FFFFFF}2    & \cellcolor[HTML]{FFFFFF}2    & \cellcolor[HTML]{FFFFFF}26\% (263) & \cellcolor[HTML]{FFFFFF}6\% (62)   & \cellcolor[HTML]{38FFF8}7\% (68)     & \cellcolor[HTML]{38FFF8}24\% (241)   & \cellcolor[HTML]{38FFF8}36\% (366)   \\ \hline
\textsc{Comp}                          & \cellcolor[HTML]{FFFFFF}$F1$& \cellcolor[HTML]{FFFFFF}2.8  & \cellcolor[HTML]{FFFFFF}2.8  & \cellcolor[HTML]{FFFFFF}57\% (570) & \cellcolor[HTML]{FFFFFF}30\% (302) & \cellcolor[HTML]{32CB00}5\% (51)     & \cellcolor[HTML]{32CB00}7\% (71)     & \cellcolor[HTML]{32CB00}0.6\% (6)      \\ \hline
\end{tabular}
}
\end{table}

Table~\ref{tab:rq1-hardflaky} shows the number of hard flaky tests among all the tests generated for each fitness function of each setup. Recall from Definition~\ref{def:sf-hf} that hard flakiness is Boolean.  As the table shows, the percentages of hard flaky tests for \textsc{PID} are between $4$\% to $16$\% for its four fitness functions. For \textsc{Pylot}, $1$\% and $6$\% of the tests are, respectively,  hard flaky for $F_1$ and $F_2$. For $F_1$ and $F_2$ of \textsc{Tran}, there are, respectively, $40$\% and $4$\% hard flaky tests. \textsc{BeamNG} yields the most hard flaky tests ranging between $22$\% to $74$\%, while for \textsc{Comp} and  its single fitness function $F_1$, we have a low hard flaky rate ($1$\%).

\begin{table}[t]
\begin{center}
\caption{Number of hard flaky tests for our ADS test setups.}
\label{tab:rq1-hardflaky}
\scalebox{1}{
\begin{tabular}{|p{0.8cm}|p{1.5cm}|p{1.5cm}|p{1.5cm}|p{1.5cm}|p{1.1cm}|}
\hline
Fitness & \textsc{PID}  & \textsc{Pylot} & \textsc{Tran} & \textsc{BeamNG} & \textsc{Comp}  \\ \hline
$F_1$ & 60 $\approx$ $6$\%  & 12  $\approx$ $1.2$\%   & 40  $\approx$ $40$\%  & 669 $\approx$ $66$\% & 9 $\approx$ 1\%      \\ \hline
$F_2$ & 66 $\approx$ $6$\%  & 64   $\approx$ $6$\%  & 4   $\approx$ $4$\%   & 744 $\approx$ $74$\%  & -   \\ \hline
$F_3$ & 163 $\approx$ $16$\%  & 0      & 0    & 328 $\approx$ $32$\%   & -   \\ \hline
$F_4$ & 40 $\approx$ $4$\%  & 0     & 0    & 227 $\approx$ $22$\% & -   \\ \hline
\end{tabular}
}
\end{center}
\vspace*{-.5cm}
\end{table}

We do not conduct further experiments with the \textsc{Tran} setup due to its prohibitive computational cost. The \textsc{Comp} setup is also excluded from the subsequent research questions due to its simple inputs and having  only one fitness function. In addition, as Tables~\ref{tab:tabrq1} and \ref{tab:rq1-hardflaky} show, \textsc{Comp}  exhibits  relatively low flakiness. We discuss the relation  between the characteristics of our test setups and flakiness in Section~\ref{sec:conclusion}.

\begin{framed}
The answer to RQ1-1 is that between 4\% and 98\% of the generated tests across our five test setups exhibit noticeable variations in their fitness values, indicating a significant presence of soft flakiness. At least one of the fitness functions of \textsc{PID}, \textsc{Pylot}, \textsc{Tran}, \textsc{BeamNG}, and \textsc{Comp} exhibit hard flaky rates of $16$\%, $6$\%, $40\%$, $74$\%, and $1$\%, respectively. 
%all of our test setups exhibit noticeable soft and hard flakiness for some fitness functions.  %These results show that the prevalence of flakiness for ADS testing is comparable to the prevalence of flakiness for code-bases. 
\end{framed}

\subsection{RQ1-2 Results} 

\begin{table}[]
\begin{center}
\caption{Inconsistencies across repeated runs of flaky tests.}
\label{tab:variations-carla}
\scalebox{0.9}{
\begin{tabular}{|p{0.8cm}|p{8cm}|p{1cm}|p{1.5cm}|}
\hline
\textbf{Setup}         & \textbf{Variation}                                                                                                                                                                                               &  \textbf{Type}  &  \textbf{Freq. (out of 40)} \\ \hline
\multirow{6}{*}{\textsc{PID}} & The ego vehicle suddenly drives to the pavement.                                                                                                                                                                 & II &  1          \\ \cline{2-4} 
                     & The ego vehicle suddenly starts to turn around itself.                                                                                                                                                           & II  &  8          \\ \cline{2-4} 
                     & Minor variations in traffic light timing impact the behavior of the ego vehicle.                                                                                                                                 & III  &  4         \\ \cline{2-4} 
                     & The streets disappear and the ego vehicle is no longer on the ground.                                                                                                                                            & I   &  1          \\ \cline{2-4} 
                     & Non-ego vehicles become unstable and impact the ego vehicle.                                                                                                                                                     & II   &  2         \\ \cline{2-4} 
                     & The ego vehicle hits other vehicles non-deterministically.                                                                                                                                                           & III    &  2       \\ \hline
\multirow{6}{*}{\textsc{Pylot}}    &  
The ego vehicle stops behind a red traffic light (or behind a vehicle that is  waiting for a red traffic light) and fails to start moving after the light turns green. 
& III            &  5                  \\ \cline{2-4} 
                     & The ego vehicle stops for no apparent reason while turning.                                                                                                                                                      & III      &  2     \\ \cline{2-4} 
                     & The ego vehicle stops for no apparent reason while driving on a straight road.                                                                                                                                   & III     &  1      \\ \cline{2-4} 
                     & The ego vehicle hits other vehicles non-deterministically.                                                                                                                                                           & III     &  7      \\ \cline{2-4} 
                     & Randomness in the ego vehicle's steering for no reason.                                                                                                                                                          & III     &  3      \\ \cline{2-4} 
                     & Randomness in the ego vehicle's behaviour when close to a non-ego vehicle.                                                                                                                                                                & III      &  4      \\ \hline
\end{tabular}
}
\end{center}
\end{table}
 To identify variations in flaky tests, this RQ requires us to inspect simulations from test reruns. To reduce discrepancies caused by potential differences between the setups and the simulators, we focus on \textsc{PID} and \textsc{Pylot}, as both employ CARLA. We randomly select, from the tests generated for RQ1-1,  $20$ tests for \textsc{PID} and $20$ tests for \textsc{Pylot}. The selection was made such that for each fitness function of each setup, we selected at least five tests that exhibit non-negligible soft flakiness (i.e., a soft flakiness ratio higher than $5\%$). Two co-authors then watched the videos of ten reruns of the selected tests to identify variations in the ego vehicle's behavior that contributed to the flakiness of the fitness values. In all the selected tests, we identified visually-visible variations in the behaviour of the ego vehicle, leading to the differences in the fitness values. 
After the initial viewing, the co-authors re-examined the videos, compared the observed variations, and summarized them in Table~\ref{tab:variations-carla}. Note that the variations describe the differences observed among multiple runs of the same test input.  For example, ``\emph{The ego vehicle hits other vehicles non-deterministically.}'' indicates  that, in some runs of a given test input the ego vehicle collides with other vehicles, while in some other runs, it does not. The videos from which the variations in Table~\ref{tab:variations-carla} are extracted are available online~\citep{github}.

% what was valid and what was obviously invalid and abnormal. 
We classify variations into three types: (I)~Infeasible scenarios that violate fundamental physics principles. (II)~Significant deviations in the ADS controller's expected behavior due to incorrect set-points, input frequencies exceeding plant bandwidth, or noise-corrupted inputs from faulty sensors (e.g., the example in Figure~\ref{fig:figflakyhard}). (III)~Scenarios that slightly differ from one another but are normal and respect both physical laws and controller behavior (e.g., the example in Figure~\ref{fig:figflaky}). Table~\ref{tab:variations-carla} presents the types of variations as well as the frequencies of the occurrence of these variation types in the analyzed videos. Among the three variation types, only scenarios with type I variations should be excluded from ADS test results since they represent flawed scenarios. Type II and III, however, 
represent meaningful scenarios and may help with revealing actual failures and with fault detection. Hence, they should not be disregarded.

\begin{framed}
The answer to RQ1-2 is that, among 40 randomly selected flaky test samples, only one sample yield flawed test reruns of type I. The reruns of the rest of the samples represent variations that  are due to unstable controllers (type II) or minor differences in the behavior of the ego vehicle (type III). These variations should not be discarded and should be further investigated to determine whether, or not, they indicate valid failures in the ADS behaviour.
\end{framed}

%Table 6 --> PID has access to the ground truth data since PID receives data from the simulator directly without having a separate component. But Pylot has a separate perception component. 

\subsection{RQ1-3 Results}  
This research question aims to evaluate the \emph{scale} of the impact of flaky tests the performance of ADS testing.
We apply the random testing algorithm (Algorithm~\ref{alg:RS}) to \textsc{PID}, \textsc{Pylot} and \textsc{BeamNG}. We run Algorithm~\ref{alg:RS} once with $n=1$ and once with $n=10$.  We refer to the former as $\mathit{RS_{n=1}}$ and to the latter as $\mathit{RS_{n=10}}$. We run each of them for 50 iterations and record the best fitness function value ($f_{\mathit{opt}}$) at each iteration. Note that $\mathit{RS_{n=1}}$ does not account for test flakiness while  $\mathit{RS_{n=10}}$ records, for each test input, the best (or lowest) fitness value obtained based on ten re-executions.  To statistically compare the results, we repeat each 50-iteration run of these algorithms 20 times.

\textbf{Metrics.} We compare $\mathit{RS_{n=1}}$ and   $\mathit{RS_{n=10}}$  using two metrics commonly used in the literature to assess ADS testing algorithms~\citep{5342440,raja2018}: (1) The number of failure revealing tests generated by each algorithm, and (2) the fitness values of the tests generated by each algorithm, where the algorithm that produces more optimal fitness values is considered better.

%The latter is determined by the fitness-function values. %In our context, a smaller value of $f_{\mathit{opt}}$ indicates a more critical failure revealed by its corresponding test case. 

\textbf{Results.}  Table~\ref{tab:tabrq11} compares the number of failure-revealing tests identified by each of $\mathit{RS_{n=10}}$ and $\mathit{RS_{n=1}}$. A test is failure-revealing with respect to a fitness function  if the test's fitness value falls  below the nominal threshold for that function (see Section~\ref{sec:background}). As Table~\ref{tab:tabrq11} shows, for all the fitness functions of \textsc{PID} and \textsc{BeamNG}, and for two fitness functions of \textsc{Pylot}, $\mathit{RS_{n=10}}$  detects more failures than $\mathit{RS_{n=1}}$. For functions $F_3$ and $F_4$ of \textsc{Pylot} neither $\mathit{RS_{n=10}}$ nor $\mathit{RS_{n=1}}$ identifies any failures. 

\begin{table}[t]
\begin{center}
\caption{Comparing the numbers of failure-revealing tests obtained by $\mathit{RS_{n=10}}$ and $\mathit{RS_{n=1}}$ for different ADS test setups.}
\label{tab:tabrq11}
\scalebox{0.95}{
\begin{tabular}{|p{1cm}|p{3cm}|p{3cm}|p{3cm}|}
\hline
Fitness & \textsc{PID}  & \textsc{Pylot}    & \textsc{BeamNG} \\ \hline
       & $\mathit{RS_{n=10}}$ - $\mathit{RS_{n=1}}$ & $\mathit{RS_{n=10}}$ - $\mathit{RS_{n=1}}$ & $\mathit{RS_{n=10}}$ - $\mathit{RS_{n=1}}$  \\ \hline
$F_1$  & 611 - 203     & 12 - 0    & 669 - 240  \\ \hline
$F_2$   & 645 - 0      & 66 - 19           & 744 - 712         \\ \hline
$F_3$    & 888 - 122          & 0 - 0              & 328 - 325    \\ \hline
$F_4$  & 489 - 225          & 0 - 0              &  227 - 129         \\ \hline
\end{tabular}
}
\end{center}
\vspace*{-.3cm}
\end{table}

%\begin{table}[t]
%\caption{Comparing the numbers of failure-revealing tests obtained by $\mathit{RS_{n=1}}$ and $\mathit{RS_{n=10}}$ for different ADS test setups.}
%\label{tab:tabrq11}
%\scalebox{0.45}{
%\begin{tabular}{|p{1cm}|p{3cm}|p{3cm}|p{3cm}|p{3cm}|p{3cm}|}
%\hline
%Fitness & \textsc{PID}  & \textsc{Pylot}              & \textsc{Tran} & \textsc{BeamNG}  &  \textsc{Comp}      \\ \hline
%        & RS(n=10) - RS(n=1) & RS(n=10) - RS(n=1) & RS(n=10) - RS(n=1) & RS(n=10) - RS(n=1) & RS(n=10) - RS(n=1)\\ \hline
%$F_1$  & 611 - 203          & 12 - 0             & 5 - 0  & 669 - 240  &  279 - 277          \\ \hline
%$F_2$   & 645 - 0            & 66 - 19            & 1 - 0 & 744 - 712  & -        \\ \hline
%$F_3$    & 888 - 122          & 0 - 0              & 50 - 50  & 328 - 325  & -     \\ \hline
%$F_4$  & 489 - 225          & 0 - 0              & 23 - 8  & 227 - 129  & -        \\ %\hline
%\end{tabular}
%}
%\end{table}

\begin{figure}[t]
    \centering
    \begin{subfigure}{\columnwidth}
      \centering
        \includegraphics[width=\columnwidth]{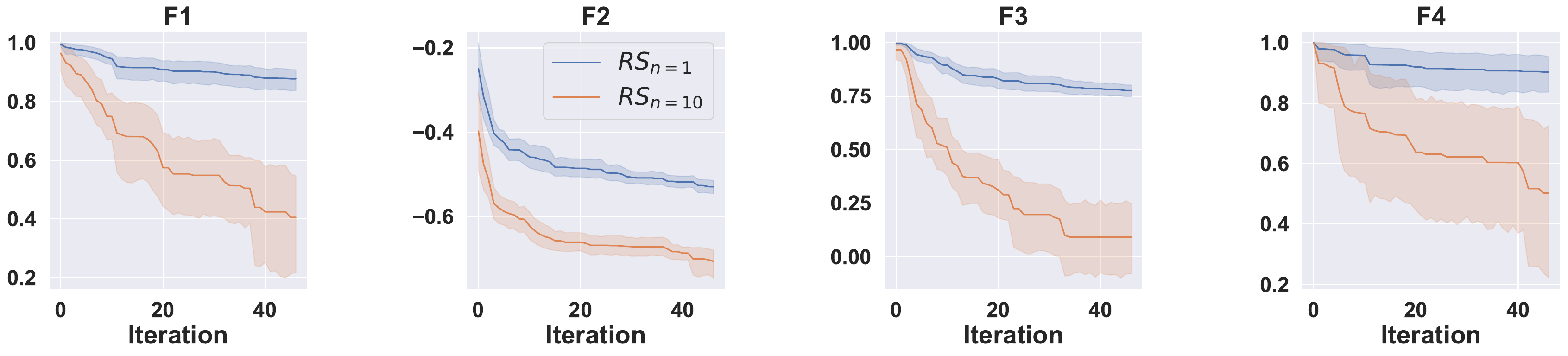}
        \caption{\textsc{PID}}
        \label{fig:rq1-carla}
        
    \end{subfigure}
    \hfil
    \begin{subfigure}{\columnwidth}
      \centering
        \includegraphics[width=\columnwidth]{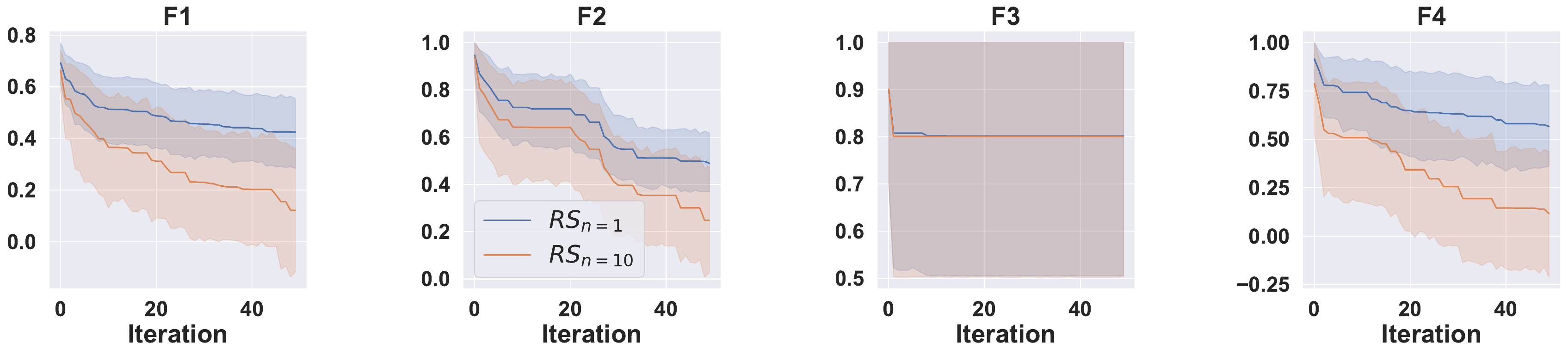}
        \caption{\textsc{Pylot}}
        \label{fig:rq1-pylot}
    \end{subfigure}
    \hfil
    \begin{subfigure}{\columnwidth}
      \centering
        \includegraphics[width=\columnwidth]{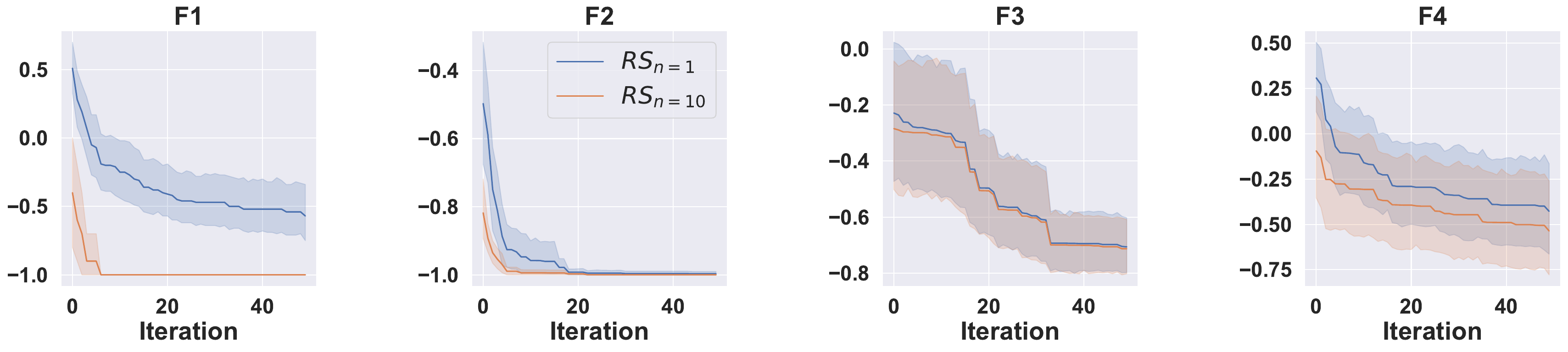}
        \caption{\textsc{BeamNG}}
        \label{fig:rq1-beamng}
    \end{subfigure}
    \caption{The averages and $95$\% intervals of the best fitness values obtained by $20$ runs of $\mathit{RS_{n=1}}$ and $\mathit{RS_{n=10}}$ over 50 iterations for four fitness functions of \textsc{PID}, \textsc{Pylot} and \textsc{BeamNG}.}
    \label{fig:rq1}
\end{figure}

 Figures~\ref{fig:rq1}(a-c) show the trends for the averages  and $95\%$ confidence intervals of best fitness values obtained from 20 runs of $\mathit{RS_{n=1}}$ and $\mathit{RS_{n=10}}$ over 50 iterations for four fitness functions of \textsc{PID}, \textsc{Pylot} and \textsc{BeamNG}, respectively. The distributions of  the final best fitness values, i.e., the fitness values at iteration $50$, obtained from $\mathit{RS_{n=1}}$ and $\mathit{RS_{n=10}}$  are available online~\citep{sup_material}. Table~\ref{tab:stattest} compares these distributions using the Wilcoxon signed-rank test~\citep{capon:91} and the Vargha Delaney $\hat{A}_{12}$ effect size~\citep{vargha}. As shown in Table~\ref{tab:stattest},  $\mathit{RS_{n=10}}$ outperforms $\mathit{RS_{n=1}}$ significantly in finding more optimal fitness values in all the cases. Further, the comparison yields a large effect size for the four fitness functions of \textsc{PID}, three fitness functions of \textsc{BeamNG}, and two fitness functions of 
\textsc{Pylot}. 

\begin{table}[t]
\begin{center}
\caption{Statistical test and effect size, Wilcoxon p-value and Vargha-Delaey $\hat{A}_{12}$ results comparing the distributions of best fitness values obtained from $\mathit{RS_{n=10}}$ and $\mathit{RS_{n=1}}$ at the last iteration of Figure~\ref{fig:rq1}.}
\label{tab:stattest}
\scalebox{1}{\begin{tabular}{|l|ll|ll|ll|}
\hline
Fitness        & \multicolumn{2}{l|}{\textsc{PID}}                        & 
        \multicolumn{2}{l|}{\textsc{Pylot}}                        & 
        \multicolumn{2}{l|}{\textsc{BeamNG}}
        \\ \hline
 & \multicolumn{1}{l|}{p-value}       & $\hat{A}_{12}$          & \multicolumn{1}{l|}{p-value}     & $\hat{A}_{12}$ & \multicolumn{1}{l|}{p-value}     & $\hat{A}_{12}$     \\ \hline
\emph{F1}      & \multicolumn{1}{l|}{7.50$e-10$} & 0.98 (L) & \multicolumn{1}{l|}{6.62$e-156$} & 0.76 (L) & \multicolumn{1}{l|}{7.53e-10} & 0.98 (L)  \\ \hline
\emph{F2}      & \multicolumn{1}{l|}{7.50$e-10$}  & 0.96 (L) & \multicolumn{1}{l|}{8.78$e-141$} & 0.76 (L) & \multicolumn{1}{l|}{7.29e-10} & 0.76(L)\\ \hline
\emph{F3}      & \multicolumn{1}{l|}{7.46$e-10$}  & 0.94 (L) & \multicolumn{1}{l|}{1.03$e-104$} & 0.62 (S) & \multicolumn{1}{l|}{7.54e-10} & 0.58 (S) \\ \hline
\emph{F4}      & \multicolumn{1}{l|}{7.50$e-10$}  & 0.89 (L) & \multicolumn{1}{l|}{6.85$e-144$} & 0.70 (M) & \multicolumn{1}{l|}{7.53e-10} & 0.79 (L)\\ \hline
\end{tabular}}
\end{center}
\vspace*{-.6cm}
\end{table}

%PLEASE REVISIT THIS PART: expressed as multiples of the standard deviation. The thresholds range from -3 to 3, with increments of 1. The results in the table highlight the importance of considering the minimum value of repeats when aiming to reach the global minimum. They also demonstrate that a greater proportion of values fall on the lower end of the spectrum. It is worth noting that, in order to provide a more accurate representation of the data, we have used a range of threshold that are multiples of the standard deviation instead of an arbitrary values, this allows for a more direct comparison of the fitness values with respect to the typical variations of the distribution.

\begin{framed}
The answer to RQ1-3 is that, for our three ADS test setups,  a random testing algorithm that  reruns ADS tests multiple times significantly outperforms an algorithm that runs ADS tests once with large effect sizes for at least two fitness functions of each test setup. In addition, the former algorithm detects significantly more failures, ranging from $12$ to $888$, and  yields  fitness values that are significantly more optimal. 
\end{framed}

% The results indicate a statistically significant difference in convergence for at least three of the four fitness metrics when considering either the minimum fitness over 10 repeats of the scenario at each iteration, the average of repeats or simply taking the first execution of the scenario.
%Figures~\ref{fig:endbox} (a-b) illustrate the distribution of our fitness metrics over the entire set of 1000 scenarios. The color coding and method of calculation are the same as in Figure~\ref{fig:rq1}. The results show that using the minimum value significantly lowers our values for three of the aforementioned metrics.

%\begin{figure}[h]
%    \centering
%    \begin{subfigure}{\columnwidth}
%        \centering
%        \includegraphics[width=\columnwidth]{count_hist}
%        \caption{CARLA + Pylot}
%        \label{fig:counthist-pylot}
%    \end{subfigure}
%    \caption{Distribution of our fitness values around their respective mean (by 3 standard deviations).}
%    \label{fig:counthist}
%\end{figure}

\subsection{RQ2-1 Results}  
\label{sec:rq2-1}
This research question evaluates if machine learning classifiers can be used to predict flaky ADS tests in an accurate and cost-effective manner. For this research question,  we develop ML classifiers to predict flaky test inputs. As discussed earlier in Section~\ref{sec:eval}, we focus on using ML to predict soft flakiness. To develop training data for predicting soft flakiness, we label a test input as flaky if it yields a soft flaky ratio higher than 5\% for at least one fitness function (see  Table~\ref{tab:tabrq1} for the soft flaky ratio results). Through additional experiments using thresholds of 10\% and 20\%, we have verified that our findings for RQ2-1 are not sensitive to the 5\% threshold  (see our supplementary material available online~\citep{sup_material}).

As discussed earlier, we explore two alternative input feature designs for our classifiers: \Ofit, which utilizes test inputs and individual fitness  values, and \Odif, which employs test inputs and the maximum difference of fitness values from multiple executions of the same test.  We evaluate four alternative subsets of test input variables to train classifiers: 1)~All input variables; 2)~Excluding ambient conditions variables; 3)~Excluding scene layout variables; 4)~Excluding both ambient conditions and scene layout variables. To build the classifiers, we investigate the following widely-used techniques: Decision Trees (DT), Random Forests (RF), Support Vector Machines (SVM), and Multi-Layer Perceptrons (MLP)~\citep{dataminingbook,hagan1997neural}.

We use the 1000 test inputs generated for \textsc{PID}, \textsc{Pylot} and \textsc{BeamNG} in RQ1-1 to develop training and test datasets for \Ofit\ and \Odif classifiers. We refer to the datasets corresponding to  \Ofit\ and \Odif\ as \Dfit\ and \Ddif, respectively. For \Dfit, we match each test input with the fitness value obtained from its first execution, resulting in one data point per test input. For \Ddif, we couple each test input with a fitness value difference derived from two or more re-executions of that test input. 
We compute nine fitness differences per test input based on the ten consecutive re-executions of each test input. After removing duplicates, \Ddif\ contains 4796 data points for \textsc{PID}, 3181 data points for \textsc{Pylot}, and 6018 for \textsc{BeamNG}.

%We choose to predict soft flakiness, rather than focusing on hard flakiness. This is because hard flakiness is a more restrictive concept, contingent upon a nominal and domain-specific threshold, and does not encompass soft flakiness. Further, as shown by RQ1-3, soft flakiness significantly impacts the results of randomized ADS testing algorithms. Hence, we focus on predicting test inputs exhibiting a non-negligible degree of soft flakiness. 

To construct each data point in \Dfit, only a single test execution is required, resulting in a consistent cost for all data points in \Dfit. In contrast, data points in \Ddif are obtained from a minimum of two to a maximum of ten re-executions, leading to varying costs associated with the data points in \Ddif. We partition \Ddif\ into nine subsets \Didif{2} to \Didif{10} where \Didif{i} contains the fitness differences obtained from the first $i$ executions.  In other words, developing \Didif{i} requires a consistent cost of $i$ executions per data point. Since developing different \Didif{i} partitions requires varying levels of effort, we assess the \Odif classifiers' performance  based on \Didif{i} datasets individually.
Note that we train \Odif classifiers on the entire \Didif\ , but for testing, we apply them to \Didif{2} to \Didif{9} separately. We do not use \Didif{10} as a test set since  the effort required to build \Didif{10} is equivalent to the effort needed for developing the ground truth (actual) labels, and the actual labels can be trivially derived for 
\Didif{10}.

Finally, we compare our classifiers with a simple, non-ML baseline that labels a data point in \Ddif flaky if the fitness difference associated to the data point is higher than the same threshold used for ground truth labeling, i.e., the threshold obtained as a soft flaky ratio of higher than $5$\%. Note that this baseline is only relevant to the \Ddif datasets since it  requires a fitness difference, and hence, at least two re-executions. We do not have any baselines, either from the literature or otherwise, that can operate on the \Dfit datasets.

%Note that, for \Didif{10}, this non-ML baseline is equal to the ground truth. But for $i <10$, this baseline likely under-performs the ground truth. While the use of \Didif{10}  is not justified for testing, we still use it for training since this set is necessary for developing  the ground truth and might as well be used for training. 

\textbf{Metrics.} We assess our classifiers using standard metrics, namely \emph{Precision}, i.e., the ability of a classifier to precisely predict flaky test inputs, \emph{Recall}, i.e., the ability of a classifier to predict all flaky test cases, and the \emph{F1-Score}, i.e., the harmonic mean of precision and recall~\citep{f1score}.  We used a 5-fold stratified cross-validation to ensure our models are trained and tested in a valid and unbiased way. For that, we allocated $80$\% of the data points for training and $20$\% for testing. Since our datasets  might be imbalanced, we use synthetic minority oversampling technique (SMOTE)~\citep{smote} for balancing the data in the training set.  We did not balance the test datasets to ensure that our model is only tested on the actual set of test inputs. 

%However, we selected the test folds such that the number of flaky and non-flaky datapoints are equal. 

%Since developing different \Didif{i}  partitions involves different degrees of effort, we report the classifiers' metrics for datasets derived from each partition separately. 

%developing \Didif{10} require highest and a bit lower than the effort required for developing the ground truth.  The input features in the \Ddif dataset include the maximum fitness differences which, as mentioned earlier, is the variable used to develop the ground truth, i.e., the actual flaky labels.  Note that \Didif{10} is the same as the dataset used for developing the ground truth label.  Hence, while we use \Didif{2} to \Didif{10} to train classifiers, for testing the classifiers, we consider \Didif{2} to \Didif{9} only. Further, the effort involved with test sets based on \Didif{2} is the least, i.e., only two executions, while 

\textbf{Results.} 
We developed and assessed $48$ classifiers ($4$ classification techniques $\times$ $4$ subsets of input variables $\times$ $3$ setups) based on \Dfit. Table~\ref{tab:rq4-bestmodels} shows the \Ofit classifiers with the highest F1-Score for each setup: For \textsc{PID}, DT with all variables; for \textsc{Pylot}, DT with all but the scene layout variables; and for \textsc{BeamNG}, RF with all but weather condition variables are the best \Ofit classifiers. In all three cases, the classifiers achieve high precision and recall values, i.e., recall more than $76$\% and precision more than $89$\%. 

\begin{table}[t]
\caption{The $\Ofit$ classifiers with the highest F1-Score}
\label{tab:rq4-bestmodels}
    \centering
\scalebox{0.9}{
\begin{tabular}{|l|l|l|l|l|l|}
\hline
\textbf{Setup} & \textbf{Inputs}                                 & \textbf{Technique} & \textbf{Precision} & \textbf{Recall} & \textbf{F1-Score} \\ \hline
\textsc{PID}          & All variables                & DT                  & 0.89               & 0.93            & 0.85              \\ \hline
\textsc{Pylot}        & All variables w/o scene layout & DT                  & 0.92               & 0.76            & 0.82              \\ \hline
\textsc{BeamNG}        & All variables w/o weather & RF                  & 1.00               & 0.93            & 0.96              \\ \hline
\end{tabular}
}
\vspace*{-.5cm}
\end{table}

Similar to the above, we developed and assessed $48$ \Odif classifiers, and selected the best for each test setup. For both \textsc{Pylot} and \textsc{PID}, MLP with all variables, and for \textsc{BeamNG}, RF with all variables yield the highest F1-Scores.  Table~\ref{tab:rq4-bestmodelsdif}  compares the precision, recall and F1-Score values of the baseline and the best \Odif classifiers obtained using test sets \Didif{2} to \Didif{9} for each setup. The second leftmost column of Table~\ref{tab:rq4-bestmodelsdif} shows the number of executions which corresponds to $i$ in \Didif{i}.

Both the baseline and the ground truth classify a datapoint as flaky when the fitness difference exceeds a specific threshold, utilizing the same threshold value. If the fitness difference exceeds the threshold after fewer than ten re-executions, it will also surpass the threshold when calculated from ten re-executions. Hence, the baseline's precision is consistently $100$\% and better than that of the \Odif classifiers. However,  the baseline's recall is lower than the classifiers' recall for most cases. In particular, since ADS testing is time consuming, we are interested in the results for low-cost test sets, i.e. those built using four or fewer executions. In Table~\ref{tab:rq4-bestmodelsdif}, we have highlighted the results for low-cost test sets in grey. \Odif classifiers notably outperform the baseline for low-cost test sets with a precision margin ranging from $5$\% to $38$\%.  Specifically, for the least expensive test set (\Didif{2}), \Odif classifiers yield F1-Score values that are $30$\%, $9$\% and $11$\% higher that those of the baseline for \textsc{PID}, \textsc{Pylot} and \textsc{BeamNG}, respectively. Full results for all the $96$ \Ofit and \Odif classifiers are available online~\citep{sup_material}.

%Note that we use the assessment over the entire \Dfit\ to select the best classifiers, but, as mentioned earlier, to compare the best classifiers against our non-ML baseline  we consider partitions \Didif{2} and \Didif{9}, i.e., the datasets obtained based two to nine  executions.

\begin{table}[t]
\caption{Comparing the best $\Odif$  classifiers with our non-ML-baseline for different ADS test setups}
\label{tab:rq4-bestmodelsdif}
    \centering
\scalebox{0.9}{
\begin{tabular}{|l|l|lll|lll|}
\hline
                                 &                                                                                   & \multicolumn{3}{l|}{\textbf{Non-ML Baseline}}                                                                                                 & \multicolumn{3}{l|}{\textbf{ML Classifier}}                                                                                                   \\ \cline{3-8} 
\multirow{-2}{*}{\textbf{Setup}} & \multirow{-2}{*}{\textbf{\begin{tabular}[c]{@{}l@{}}\# of \\ Execs\end{tabular}}} & \multicolumn{1}{l|}{\textbf{Precision}}           & \multicolumn{1}{l|}{\textbf{Recall}}              & \textbf{F1-Score}                     & \multicolumn{1}{l|}{\textbf{Precision}}           & \multicolumn{1}{l|}{\textbf{Recall}}              & \textbf{F1-Score}                     \\ \hline
                                 & 2                                                                                 & \multicolumn{1}{l|}{\cellcolor[HTML]{C0C0C0}1.00} & \multicolumn{1}{l|}{\cellcolor[HTML]{C0C0C0}0.37} & \cellcolor[HTML]{C0C0C0}0.54 & \multicolumn{1}{l|}{\cellcolor[HTML]{C0C0C0}0.97} & \multicolumn{1}{l|}{\cellcolor[HTML]{C0C0C0}0.75} & \cellcolor[HTML]{C0C0C0}\textbf{0.84} \\ \cline{2-8} 
                                 & 3                                                                                 & \multicolumn{1}{l|}{\cellcolor[HTML]{C0C0C0}1.00} & \multicolumn{1}{l|}{\cellcolor[HTML]{C0C0C0}0.53} & \cellcolor[HTML]{C0C0C0}0.70 & \multicolumn{1}{l|}{\cellcolor[HTML]{C0C0C0}0.97} & \multicolumn{1}{l|}{\cellcolor[HTML]{C0C0C0}0.86} & \cellcolor[HTML]{C0C0C0}\textbf{0.91} \\ \cline{2-8} 
                                 & 4                                                                                 & \multicolumn{1}{l|}{\cellcolor[HTML]{C0C0C0}1.00} & \multicolumn{1}{l|}{\cellcolor[HTML]{C0C0C0}0.66} & \cellcolor[HTML]{C0C0C0}0.79 & \multicolumn{1}{l|}{\cellcolor[HTML]{C0C0C0}0.97} & \multicolumn{1}{l|}{\cellcolor[HTML]{C0C0C0}0.91} & \cellcolor[HTML]{C0C0C0}\textbf{0.94} \\ \cline{2-8} 
                                 & 5                                                                                 & \multicolumn{1}{l|}{1.00}                         & \multicolumn{1}{l|}{0.75}                         & 0.85                                  & \multicolumn{1}{l|}{0.96}                         & \multicolumn{1}{l|}{0.93}                         & \textbf{0.95}                                  \\ \cline{2-8} 
                                 & 6                                                                                 & \multicolumn{1}{l|}{1.00}                         & \multicolumn{1}{l|}{0.82}                         & 0.90                                  & \multicolumn{1}{l|}{0.96}                         & \multicolumn{1}{l|}{0.96}                         & \textbf{0.96}                                  \\ \cline{2-8} 
                                 & 7                                                                                 & \multicolumn{1}{l|}{1.00}                         & \multicolumn{1}{l|}{0.87}                         & 0.93                                  & \multicolumn{1}{l|}{0.96}                         & \multicolumn{1}{l|}{0.96}                         & \textbf{0.96}                                  \\ \cline{2-8} 
                                 & 8                                                                                 & \multicolumn{1}{l|}{1.00}                         & \multicolumn{1}{l|}{0.92}                         & 0.96                                  & \multicolumn{1}{l|}{0.96}                         & \multicolumn{1}{l|}{0.98}                         & \textbf{0.97}                                  \\ \cline{2-8} 
\multirow{-8}{*}{\textsc{PID}}            & 9                                                                                 & \multicolumn{1}{l|}{1.00}                         & \multicolumn{1}{l|}{0.95}                         & \textbf{0.97}                                  & \multicolumn{1}{l|}{0.95}                         & \multicolumn{1}{l|}{0.99}                         & \textbf{0.97}                                  \\ \hline
                                 & 2                                                                                 & \multicolumn{1}{l|}{\cellcolor[HTML]{C0C0C0}1.00} & \multicolumn{1}{l|}{\cellcolor[HTML]{C0C0C0}0.44} & \cellcolor[HTML]{C0C0C0}0.61 & \multicolumn{1}{l|}{\cellcolor[HTML]{C0C0C0}0.98} & \multicolumn{1}{l|}{\cellcolor[HTML]{C0C0C0}0.54} & \cellcolor[HTML]{C0C0C0}\textbf{0.70} \\ \cline{2-8} 
                                 & 3                                                                                 & \multicolumn{1}{l|}{\cellcolor[HTML]{C0C0C0}1.00} & \multicolumn{1}{l|}{\cellcolor[HTML]{C0C0C0}0.60} & \cellcolor[HTML]{C0C0C0}0.75 & \multicolumn{1}{l|}{\cellcolor[HTML]{C0C0C0}0.96} & \multicolumn{1}{l|}{\cellcolor[HTML]{C0C0C0}0.73} & \cellcolor[HTML]{C0C0C0}\textbf{0.83} \\ \cline{2-8} 
                                 & 4                                                                                 & \multicolumn{1}{l|}{\cellcolor[HTML]{C0C0C0}1.00}  & \multicolumn{1}{l|}{\cellcolor[HTML]{C0C0C0}0.70} & \cellcolor[HTML]{C0C0C0}0.82 & \multicolumn{1}{l|}{\cellcolor[HTML]{C0C0C0}0.95} & \multicolumn{1}{l|}{\cellcolor[HTML]{C0C0C0}0.83} & \cellcolor[HTML]{C0C0C0}\textbf{0.89} \\ \cline{2-8} 
                                 & 5                                                                                 & \multicolumn{1}{l|}{1.00}                          & \multicolumn{1}{l|}{0.77}                         & 0.87                                  & \multicolumn{1}{l|}{0.94}                         & \multicolumn{1}{l|}{0.88}                         & \textbf{0.91}                                  \\ \cline{2-8} 
                                 & 6                                                                                 & \multicolumn{1}{l|}{1.00}                          & \multicolumn{1}{l|}{0.84}                         & 0.91                                  & \multicolumn{1}{l|}{0.93}                         & \multicolumn{1}{l|}{0.92}                         & \textbf{0.92}                                  \\ \cline{2-8} 
                                 & 7                                                                                 & \multicolumn{1}{l|}{1.00}                          & \multicolumn{1}{l|}{0.90}                         & \textbf{0.94}                                  & \multicolumn{1}{l|}{0.90}                         & \multicolumn{1}{l|}{0.94}                         & 0.92                                  \\ \cline{2-8} 
                                 & 8                                                                                 & \multicolumn{1}{l|}{1.00}                          & \multicolumn{1}{l|}{0.92}                         & \textbf{0.96}                                  & \multicolumn{1}{l|}{0.89}                         & \multicolumn{1}{l|}{0.95}                         & 0.92                                  \\ \cline{2-8} 
\multirow{-8}{*}{\textsc{Pylot}}          & 9                                                                                 & \multicolumn{1}{l|}{1.00}                          & \multicolumn{1}{l|}{0.96}                         & \textbf{0.98}                                  & \multicolumn{1}{l|}{0.88}                         & \multicolumn{1}{l|}{0.97}                         & 0.92                                  \\ \hline
& 2                                                                                 & \multicolumn{1}{l|}{\cellcolor[HTML]{C0C0C0}1.00} & \multicolumn{1}{l|}{\cellcolor[HTML]{C0C0C0}0.71} & \cellcolor[HTML]{C0C0C0}0.83 & \multicolumn{1}{l|}{\cellcolor[HTML]{C0C0C0}1.00} & \multicolumn{1}{l|}{\cellcolor[HTML]{C0C0C0}0.89} & \cellcolor[HTML]{C0C0C0}\textbf{0.94} \\ \cline{2-8} 
                                 & 3                                                                                 & \multicolumn{1}{l|}{\cellcolor[HTML]{C0C0C0}1.00} & \multicolumn{1}{l|}{\cellcolor[HTML]{C0C0C0}0.86} & \cellcolor[HTML]{C0C0C0}0.92 & \multicolumn{1}{l|}{\cellcolor[HTML]{C0C0C0}0.99} & \multicolumn{1}{l|}{\cellcolor[HTML]{C0C0C0}0.93} & \cellcolor[HTML]{C0C0C0}\textbf{0.96} \\ \cline{2-8} 
                                 & 4                                                                                 & \multicolumn{1}{l|}{\cellcolor[HTML]{C0C0C0}1.00}  & \multicolumn{1}{l|}{\cellcolor[HTML]{C0C0C0}0.90} & \cellcolor[HTML]{C0C0C0}0.95 & \multicolumn{1}{l|}{\cellcolor[HTML]{C0C0C0}0.99} & \multicolumn{1}{l|}{\cellcolor[HTML]{C0C0C0}0.95} & \cellcolor[HTML]{C0C0C0}\textbf{0.97} \\ \cline{2-8} 
                                 & 5                                                                                 & \multicolumn{1}{l|}{1.00}                          & \multicolumn{1}{l|}{0.93}                         & 0.96                                  & \multicolumn{1}{l|}{1.00}                         & \multicolumn{1}{l|}{0.96}                         & \textbf{0.98}                                  \\ \cline{2-8} 
                                 & 6                                                                                 & \multicolumn{1}{l|}{1.00}                          & \multicolumn{1}{l|}{0.95}                         & 0.97                                  & \multicolumn{1}{l|}{1.00}                         & \multicolumn{1}{l|}{0.96}                         & \textbf{0.98}                                  \\ \cline{2-8} 
                                 & 7                                                                                 & \multicolumn{1}{l|}{1.00}                          & \multicolumn{1}{l|}{0.97}                         & \textbf{0.98}                                  & \multicolumn{1}{l|}{0.99}                         & \multicolumn{1}{l|}{0.97}                         & \textbf{0.98}                                 \\ \cline{2-8} 
                                 & 8                                                                                 & \multicolumn{1}{l|}{1.00}                          & \multicolumn{1}{l|}{0.98}                         & \textbf{0.99}                                  & \multicolumn{1}{l|}{0.99}                         & \multicolumn{1}{l|}{0.98}                         & \textbf{0.99}                                  \\ \cline{2-8} 
\multirow{-8}{*}{\textsc{BeamNG}}          & 9                                                                                 & \multicolumn{1}{l|}{1.00}                          & \multicolumn{1}{l|}{0.99}                         & \textbf{0.99}                                   & \multicolumn{1}{l|}{0.99}                         & \multicolumn{1}{l|}{0.98}                         & \textbf{0.99}                                   \\ \hline
\end{tabular}
}
\vspace*{-.35cm}
\end{table}

\begin{framed}
The answer to RQ2-1 is that  ML classifiers effectively identify flaky ADS test inputs with high precision and recall. Single-test-execution-based (\Ofit) classifiers  achieve at least 89\% precision and 76\% recall across various test setups. Multi-test-execution-based (\Odif) classifiers, when applied to datasets requiring four or fewer test reruns,  yield 95\% or higher precision and 83\% or higher recall, outperforming a non-ML baseline.
\end{framed}

\subsection{RQ2-2 Results}  
This research question investigates if machine learning classifiers can improve the performance of ADS testing by requiring a limited number of test reruns.
For this research question, we evaluate a random testing algorithm, denoted by $\mathit{RS_{ML}}$, that uses \Ofit and \Odif classifiers according to Algorithm~\ref{alg:smartfitnessn} to compute optimal fitness values with minimal test reruns.  The algorithm uses \Ofit predictions after the first test execution and \Odif predictions after the subsequent test executions to determine whether, or not, a test is flaky and if the test re-executions should continue. We develop a baseline for $\mathit{RS_{ML}}$, denoted  by $\mathit{RS_{b}}$, using  the non-ML baseline from RQ2-1. Recall that the non-ML baseline could only be developed based on the  \Ddif datasets as we have no baseline that could operate based on the \Dfit datasets. The $\mathit{RS_{b}}$ algorithm replaces \Ofit in line~6 of Algorithm~\ref{alg:smartfitnessn} with  ``true'' as we have no alternative for \Ofit, and replaces \Odif in line~15 of Algorithm~\ref{alg:smartfitnessn} with the non-ML baseline.

%that starts with two test executions, and repeats the test executions until the non-ML baseline of RQ2-1 keeps classifying the test as flaky or until $n$ iterations are reached.

%the non-ML baseline of RQ2-1 and is defined as follows: In line 6 of Algorithm \ref{alg:smartfitnessn}, instead of calling \Ofit, it sets \textit{flaky}  to True. In line 15, the baseline compares the difference of the maximum and the minimum fitness values with a threshold, as in RQ2-1 non-ML baseline.

%Specifically, we call \textit{fastFitness}  (Algorithm~\ref{alg:smartfitnessn}) in lines 4 and 7 of Algorithm \ref{alg:RS} for fitness calculation. 

%We compare the results with three other methods, i.e., RS(n=1), RS(n=10), and RS(non-ML).  For RS(n=1) and RS(n=10), we use the results from RQ1-3.  

We execute both $\mathit{RS_{ML}}$ and $\mathit{RS_{b}}$  for $50$ iterations and for $20$ times. We then compare their performance in obtaining optimal fitness values, while also recording the number of simulations (i.e., test executions) each algorithm performs. We set the maximum iterations for both $\mathit{RS_{ML}}$ and $\mathit{RS_{b}}$ to ten to be consistent with the $\mathit{RS}_{n=10}$ algorithm used for RQ1-3. In RQ1-3, $\mathit{RS}_{n=10}$, which outperformed $\mathit{RS}_{n=1}$, executed $10,000$ simulations ($50$ iterations $\times$ $20$ runs $\times$ $10$ re-executions). We expect $\mathit{RS_{ML}}$ and $\mathit{RS_{b}}$ to obtain fitness values close to those obtained by $\mathit{RS}_{n=10}$ but using significantly fewer simulations.

%\emph{Metrics.} (1) RS(n=1) and RS(n=10) show the extreme cases. RS(n=10) outperofms RS(n=1) while  RS(n=1) performs XX executions and RS(n=10) performs XX executions. We expect RS(ML) and RS(non-ML) to be in between. We want to determine which one provides a better trade-off between accuracy and cost. We meausre accuracy by the value of fitness function at the end (similar to figure ?? in RQ1-3). A better trade-off is the the one with accuracy closer to RS(n=10) while cost is closert to RS(n=1). 

The distributions of the best fitness values computed by $\mathit{RS_{ML}}$ and $\mathit{RS_{b}}$ are available online \citep{github}. Both $\mathit{RS_{ML}}$ and $\mathit{RS_{b}}$ obtain fitness values that are significantly more optimal than those obtained by $\mathit{RS}_{n=1}$, but less optimal than those obtained by $\mathit{RS}_{n=10}$. To compare $\mathit{RS_{ML}}$ and $\mathit{RS_{b}}$, we provide the number of simulations each algorithm performs in Table~\ref{tab:rq2-efforts}. In addition, Table~\ref{tab:rq2-2-stattest-baseline} statistically compares the distributions of fitness values obtained by $\mathit{RS_{ML}}$ and $\mathit{RS_{b}}$ for the four fitness functions of our three test setups. The results in Table~\ref{tab:rq2-2-stattest-baseline} shows that for eight fitness functions, 
$\mathit{RS_{ML}}$ obtains significantly better fitness functions than $\mathit{RS_{b}}$, while for four fitness functions,  $\mathit{RS_{b}}$ outperforms  $\mathit{RS_{ML}}$. Overall and as shown in the fitness distributions available online~\citep{sup_material}, both algorithms obtain optimal fitness values that are quite close. However, as Table~\ref{tab:rq2-efforts} shows, overall, $\mathit{RS_{ML}}$ requires much fewer simulations to compute these optimal fitness values. Specifically, $\mathit{RS_{ML}}$ requires $903$ less simulations for \textsc{PID} and $3760$ less simulations for \textsc{Pylot}. Although $\mathit{RS_{ML}}$ requires more simulations than $\mathit{RS_{b}}$ for \textsc{BeamNG}, the difference is relatively minor (only $243$ additional simulations).

\begin{table}[t]
    \centering
\caption{Number of simulations performed by $\mathit{RS_{ML}}$ and $\mathit{RS_{b}}$ for each test setup}
\label{tab:rq2-efforts}
\scalebox{1}{
\begin{tabular}{|l|l|l|l|l|l|l|l|l|}
\hline
  & $\mathit{RS_{b}}$ & $\mathit{RS_{ML}}$ &  & $\mathit{RS_{b}}$ & $\mathit{RS_{ML}}$ &   & $\mathit{RS_{b}}$ & $\mathit{RS_{ML}}$ \\ \hline
\textsc{PID}          & 5323              & 4420           & \textsc{Pylot}        & 7391              & 3631    &  \textsc{BeamNG}       & 4153              & 4396       \\ \hline
\end{tabular}
}
\end{table}

\begin{table}[t]
    \centering
\caption{Statistical tests, Wilcoxon p-value and Vargha-Delaey $\hat{A}_{12}$, comparing the best fitness values obtained by $\mathit{RS_{ML}}$ (abbreviated as ML) and $\mathit{RS_{b}}$ (abbreviate as b) after performing the number of simulations in Table \ref{tab:rq2-efforts}}
\label{tab:rq2-2-stattest-baseline}
\scalebox{0.83}{
\begin{tabular}{l|llll|llll|}
\cline{2-9}
                                   & \multicolumn{4}{c|}{p-value}                                                                                                                                                                                                                                                                                                                            & \multicolumn{4}{c|}{\textbf{$A_{12}$}}                                                                                           \\ \hline
\multicolumn{1}{|l|}{Setup} & \multicolumn{1}{l|}{\textbf{$F_1$}}                                                            & \multicolumn{1}{l|}{\textbf{$F_2$}}                                                            & \multicolumn{1}{l|}{\textbf{$F_3$}}                                                            & \textbf{$F_4$}                                                            & \multicolumn{1}{l|}{\textbf{$F_1$}}   & \multicolumn{1}{l|}{\textbf{$F_2$}}  & \multicolumn{1}{l|}{\textbf{$F_3$}}   & \textbf{$F_4$}   \\ \hline
\multicolumn{1}{|l|}{\textsc{PID}}          & \multicolumn{1}{l|}{\begin{tabular}[c]{@{}l@{}}1.23e-08 \\ (b \textgreater ML)\end{tabular}}    & \multicolumn{1}{l|}{\begin{tabular}[c]{@{}l@{}}2.37e-09 \\ (ML \textgreater b)\end{tabular}} & \multicolumn{1}{l|}{\begin{tabular}[c]{@{}l@{}}0.0003 \\ (ML \textgreater b)\end{tabular}}   & \begin{tabular}[c]{@{}l@{}}7.18e-09 \\ (b \textgreater ML)\end{tabular}    & \multicolumn{1}{l|}{0.29 (M)} & \multicolumn{1}{l|}{0.78 (L)} & \multicolumn{1}{l|}{0.54 (S)}  & 0.31 (M) \\ \hline
\multicolumn{1}{|l|}{\textsc{Pylot}}        & \multicolumn{1}{l|}{\begin{tabular}[c]{@{}l@{}}8.00e-10 \\ (ML \textgreater b)\end{tabular}} & \multicolumn{1}{l|}{\begin{tabular}[c]{@{}l@{}}1.05e-09 \\ (ML \textgreater b)\end{tabular}} & \multicolumn{1}{l|}{\begin{tabular}[c]{@{}l@{}}0.0001 \\ (ML \textgreater b)\end{tabular}}   & \begin{tabular}[c]{@{}l@{}}7.12e-10 \\ (b \textgreater ML)\end{tabular}    & \multicolumn{1}{l|}{0.85 (L)}  & \multicolumn{1}{l|}{0.71 (L)} & \multicolumn{1}{l|}{0.66 (M)} & 0.15 (L)  \\ \hline
\multicolumn{1}{|l|}{\textsc{BeamNG}}       & \multicolumn{1}{l|}{\begin{tabular}[c]{@{}l@{}}2.10e-07 \\ (b \textgreater ML)\end{tabular}}    & \multicolumn{1}{l|}{\begin{tabular}[c]{@{}l@{}}7.46e-10 \\ (ML \textgreater b)\end{tabular}} & \multicolumn{1}{l|}{\begin{tabular}[c]{@{}l@{}}7.51e-10 \\ (ML \textgreater b)\end{tabular}} & \begin{tabular}[c]{@{}l@{}}7.54e-10 \\ (ML \textgreater b)\end{tabular} & \multicolumn{1}{l|}{0.11 (L)}  & \multicolumn{1}{l|}{0.88 (L)} & \multicolumn{1}{l|}{0.59 (S)}  & 0.59 (S)  \\ \hline
\end{tabular}
}
\end{table}

\begin{framed}
The answer to RQ2-2 is that our ADS random testing algorithm that uses ML classifiers achieves fitness values that are comparable to those of a non-ML baseline, but with significantly fewer simulations required.
\end{framed}

\subsection{Threats to Validity} 
\label{sec:threats}
The most important threats concerning the validity of our experiments are related to the internal, construct and external validity.

\subsection{Internal Validity}
To mitigate \textit{internal validity} risks, which refer to confounding factors, we ensure  that the  \textsc{Pylot}, \textsc{Tran}, \textsc{PID}, and \textsc{BeamNG} test setups share the same test input space and use consistent fitness functions. Our \textsc{Comp} setup uses the same test input space as the SBFT competition benchmark~\citep{sbftgithub} and maintains consistency with the benchmark by employing a single fitness function. We normalize the ranges of the fitness functions for all the setups before computing the soft flaky metric. In addition, we have used consistent thresholds for the fitness functions to identify failures and compute hard flakiness. In particular, for the lane keeping fitness function, $F_1$, we assume a lane invasion occurs whenever the ego car's distance to the center of lane increases $0.5$ meters. For  $F_2$ and $F_3$, we assume a collision occurs when the ego car's distance to other cars or static objects decreases $0.5$ meters. For $F_4$, we assume that the ego car fails to reach its end goal when at the end of the simulation, the ego car is more than one meter away from its final destination. These thresholds are consistent with those used by the original studies~\citep{samota,haq2022manyobjective,sbftgithub}.

For our experiments, we minimize, to the best of our abilities, internal randomness factors of CARLA and BeamNG that could be controlled through their random seeds or their configuration parameters. We identify potential sources of randomness in both, as per their official documentation~\citep{carlafoundations,carlaquickstart,beamng}. Below, we outline the measures taken to control or mitigate these random elements. \\
In CARLA, we identify the traffic lights' function, traffic manager behaviour, and non-ego vehicles' behaviours as potential sources of randomness. We control traffic lights' behavior by fixing their initial states, and  controlling the duration of  red, green or yellow using a constant seed. As for the traffic manager, we set its random seed to a constant and ensure that it runs in deterministic mode by setting a the Boolean parameter that controls its mode of execution.  In CARLA, each non-ego vehicle takes a random blueprint (i.e., type, shape, and size) when it is spawned. We fix the blueprint for each non-ego vehicle in our test inputs. \\
BeamNG does not provide any API to manipulate the blueprints of non-ego vehicles and the traffic lights. Further, its AI-based ADS does not react to traffic lights. Both BeamNG and CARLA may be executed in the synchronous or asynchronous modes. To reduce randomness, we use both in their synchronous mode where the client controls the simulator's updates. In contrast, in the asynchronous mode, the simulators run as fast as possible and handles client requests on the fly.

\subsection{Construct validity}
\textit{Construct validity} threats relate to the inappropriate use of metrics. Our hard flakiness metric is consistent with the notion of flaky tests for software code~\citep{deflaker, survey}.  For soft flakiness, the key attribute a metric should possess is the ability to capture the degree of variations in fitness function values across multiple re-runs. Our current choice for the SF metric, i.e, the difference between max and min, maintains the metric's scale consistent with that of the fitness function, facilitating easier interpretation.
An alternative way to measure the SF metric is to use standard deviation. However, standard deviation presents a challenge due to its squared scaling of the fitness function values. In other words, if the differences between the fitness values and the mean exceed one, the standard deviation will be greater than the mean of these differences; and dually, if these differences are less than one, the standard deviation will be smaller.  This scaling inconsistency complicates the interpretation of an SF metric measured by standard deviation, as the metric values do not align with the scale of fitness values.

\subsection{External validity}
\textit{External validity} is related to the generalizability of our results.  The choice of our test setups  may influence the generalizability of our results. Related to this threat, we note that we strive to diversify the setups used in our study in terms of the simulators, the ADS controllers and the test input designs. Specifically, our test setups are based on two widely-used  simulators, four ADS controllers with different internal designs, and two different test input designs. Second, three of our test setups are adopted from the literature and have been previously used in ADS testing research~\citep{gog2021pylot, transfuser, sbftgithub}. Third, our aim is not to conclude a high frequency of  flakiness for all ADS testing setups, but to study the prevalence of flaky tests in ADS simulation-based testing, their impact on test results of ADS  and ways to mitigate this impact. In particular, we believe that our proposed approach for assessing the prevalence, impact, and mitigation of flakiness in ADS testing maintains relevance and applicability across a wide range of simulators in the fields of robotics and cyber-physical systems.
The above being said, our work would benefit from further experiments with a broader class of ADS test setups.

\section{Discussion}
\label{sec:conclusion}
In this section, we outline three key lessons derived from our findings in Tables~\ref{tab:tabrq1} and ~\ref{tab:rq1-hardflaky} of RQ1-1. To better illustrate the connection between the lessons and our findings, we have summarized the hard and soft flaky rate ranges from Tables~\ref{tab:tabrq1} and ~\ref{tab:rq1-hardflaky} in Table~\ref{tab:finalres}. Note that  Table~\ref{tab:finalres} shows ranges for non-negligible soft flaky rates, i.e.,  exceeding $5$\%.  In addition to the three lessons, we present two clarifications concerning the scope of our study and its impact on prior research. Finally, we share observations on how our findings could influence future research in the field of simulation-based testing of ADS.

%We investigated the prevalence of flaky tests in ADS simulation-based testing and their impact on the test results. Further, we studied the effectiveness of machine learning classifiers in predicting flaky tests cost-effectively, and their potential to mitigate the impact of flaky ADS simulators on test results through minimal test reruns. We conclude the paper withe following observations: 

\begin{table}[t]
\begin{center}
\caption{The hard and soft flaky rate ranges for our five test setups summarising the results  from Tables~\ref{tab:tabrq1} and ~\ref{tab:rq1-hardflaky}. The table considers  soft flaky rates that exceed $5$\%,  as values below this threshold are deemed negligible.} 
\label{tab:finalres}
\scalebox{1}{
\begin{tabular}{|p{1.5cm}|p{4cm}|p{4cm}|}
\hline
\textsc{Setup ID} & \textsc{Hard Flaky rate ranges} & \textsc{Soft flaky rate ranges (more that $5$\%)} \\ \hline

\textsc{PID} & $4$\% --- $16$\% & $5$\% -- $65$\%\\ \hline

\textsc{Pylot} & $0$\% --- $6$\%  & $4$\% -- $23$\% \\ \hline

\textsc{Tran} & $0$\% ---  $40$\% &  $4$\% ---  $94$\% \\ \hline

\textsc{BeamNG} & $22$\% --- $66$\%  & $47$\% ---  $68$\% \\ \hline

\textsc{Comp} & $1$\%   & $12$\% \\ \hline

\end{tabular}
}
\end{center}
\vspace*{-.5cm}
\end{table}

\emph{Lesson~1: The lane-keeping test setup shows the lowest rate of flaky tests.} The lane-keeping test setup, i.e., the \textsc{Comp} setup, consists of only the ego vehicle driving on a single-lane road and is focused on assessing the lane-keeping requirement, i.e., the \emph{R1} requirement. As shown in Table~\ref{tab:finalres}, this setup has the lowest rate of flaky tests.  Recall from Table~\ref{tab:expinfo} that the \textsc{Comp} setup shares the same ADS controller and the same simulator with the \textsc{BeamNG} setup. However, \textsc{BeamNG} and our three other setups feature a more extensive set of input variables. They encompass a map of a small town with a two-lane road intersection where other vehicles travel alongside the ego car, with several static objects also present in the environment. This richer test environment allows us to assess collision requirements, \emph{R2} and \emph{R3},  in addition to the lane keeping requirement. However,  as Table~\ref{tab:finalres} shows the restricted environment  of the \textsc{Comp} setup leads to a considerably lower flaky test rate compared to the others.

\emph{Lesson~2: Modular DNN-based ADS  reduces flaky tests compared to end-to-end DNN-based ADS.} 
As shown in Table~\ref{tab:finalres}, between the two CARLA-based setups with DNN-based ADS, i.e., \textsc{Pylot} and \textsc{Tran},  \textsc{Pylot} yields lower flaky test rates.  Given the  test inputs and outputs consistency between these two setups and their shared simulator,  we attribute the differences in their flaky rates to their ADS controller design. Across different re-executions, the simulator may pass slightly different images  to the ADS due to synchronization and timing inconsistencies or due to the white noise addition function of the simulator.  
Both \textsc{Pylot}  and \textsc{Tran} are DNN-enabled. But 
\textsc{Tran} uses a single end-to-end transformer-based network, lacking \textsc{Pylot}'s modular structure. 
\textsc{Tran}'s outputs are produced by a single DNN, making the outputs potentially sensitive to minor input image variations. In contrast, \textsc{Pylot} uses multiple DNNs, and also leverages an independent classical controller, i.e. MPC and PID~\citep{gog2021pylot}. As a result,  potential inaccuracies and noise in DNN outputs may be corrected  by the  controller, reducing the flaky test rate for the \textsc{Pylot} setup with a modular DNN-based ADS.

\emph{Lesson~3: The Carla simulator yields a lower flaky test  rate compared to the BeamNG simulator.} The \textsc{PID} and \textsc{BeamNG} setups use their respective simulators' autopilots  as ADS and have consistent inputs and outputs. Their difference is that \textsc{PID} is based on the Carla simulator, while \textsc{BeamNG} is based on the BeamNG simulator. The considerably higher flaky test rate of the \textsc{BeamNG} setup in Table~\ref{tab:finalres} suggests that, provided with the same ADS and the same test input environment, the BeamNG simulator is more prone to producing flaky outputs compared to the Carla simulator.

\emph{Clarification of the scope: Identifying the root-causes of flakiness for simulators is out of the scope of our study.} Our research is focused on studying the presence and prevalence of non-determinism in ADS simulators and assessing the impact of this non-determinism on randomized  ADS testing methods. More research is needed to understand flakiness causes in ADS simulators. Possible causes of flakiness include, among others,  simulator bugs, simulator's autopilot instability, inconsistent timing and synchronization between the ADS and the simulator, as well as uncertainties in simulator's physical models. We are not able to assert which of these issues were the root causes of flakiness in the test setups discussed in this paper. 
However, in the case of our deep learning-based ADS models, i.e., Pylot and Transfuser, we have not detected  any non-determinism during the inference phase. Hence, we believe that, in our experiments, the ADS itself is not a contributor to the observed flakiness. In other words, the DNN-based ADS under study in this paper, when used as a standalone component for inference purposes, are deterministic.
We note that some root-causes of flakiness may be inevitable in this domain. Assuming that the simulator is bug-free, we have noticed that these issues may still cause non-determinism: (1)~inconsistent timing and synchronization between the ADS and  the simulator. Across different re-executions, the simulator may pass slightly different images or slightly different sensoray measurements to the ADS due to synchronization and timing inconsistencies. (2)~Uncertainties in the simulatoin environment due to the presence of non-ego cars, traffic lights and pedestrians, etc. (3)~White noise addition. Adding white noise to images is a common practice in data augmentation for machine learning and simulations, mimicking real-world disturbances~\citep{carlafoundations}. If the simulator's images are used for training, the white noise addition helps robustify the ML models and improve their generalizability. 
In our experiments, we noticed that both Carla and BeamNG simulators add white noise to each frame.
The white noise added by the simulator, however, may contribute to flakiness when we use simulators for testing. 

Based on our observations, the root-causes of flakiness are related to the simulator or to the interactions between the ADS and the simulator. Hence, the flakiness predictors may not generalize beyond a simulator or even a pair of simulator and ADS.  Therefore, in Section~\ref{sec:rq2-1},  we train a predictor for each pair of simulator and ADS. %Hence,  in Section~\ref{sec:rq2-1}, we train predictors for each pair of simulator and ADS with inputs invariant of ADS. 
%This allows the developers that use the same simulator and ADS pair, e.g. Carla with Pylot, use our pre-trained flakiness predictors in order to optimize the number of reruns for each scenario.

In RQ1-2, we identified three different types of variations in the flaky simulations that we checked manually. Based on our observations, the type I and II variations can be attributed to the simulator,  while type III variations are likely a result of synchronization timing inconsistencies between the simulator and the ADS or are due to the white noise addition. This observation is indeed consistent with the results shown in Table~\ref{tab:variations-carla}. Specifically, Table~\ref{tab:variations-carla} shows that the \textsc{PID} setup, which uses the simulator's PID controller as ADS, exhibits the observed type I and II variations. This supports our hypothesis that the simulator’s autopilot, specifically Carla PID, is the source of these inconsistencies.  Conversely,  most type III inconsistencies are related to the \textsc{Pylot} setup, which uses an external DNN-based ADS, and is prone to issues caused by the timing inconsistencies between the ADS and the simulator and the white noise addition.

\emph{Clarification of the impact: Our study does not invalidate the comparison results of prior simulation-based ADS testing research, but the absolute values of metrics in these studies may be impacted by flakiness.} Previous research on simulation-based ADS testing typically involves re-running proposed test strategies for different sampled inputs and conducting statistical comparisons. For studies focused on comparing different test strategies, this approach accounts for the test strategy's randomness and the potential non-determinism in ADS simulators. While the relative comparison of metrics remains unaffected, the absolute values of metrics, such as the number of failure scenarios and optimal fitness values, can be impacted by simulator flakiness. This indicates a need to revisit the metrics and assessment methods for ADS testing research, which we will discuss subsequently.

\emph{Implications of our results for future research on simulation-based ADS testing:} Considering the engineering effort needed to set up a simulator and integrate it with an ADS for testing purposes, research papers should place more emphasis on clearly detailing and characterizing the integration process and test input specifications. Currently, greater focus is placed on devising different heuristics for test strategies, while ADS simulators are often treated as black boxes. Given a test input design, the degree of potential randomness and non-determinism incurred by a simulator used for testing should be explicitly measured and reported.

To perform ADS testing, we are faced with a spectrum of possibilities for configuring ADS test setups. On one extreme, we may consider a complex urban map with an arbitrary number of non-ego vehicles and pedestrians. This setup, while allowing us to test ADS for a variety of safety requirements and situations, likely leads to a significant flaky test rate. On another extreme, we may consider a restricted map with no, or few, fixed-behavior and controlled mobile objects other than the ego vehicle. This setup, while being restricted, likely has a low or negligible flaky test rate. While fitness values of individual scenarios can be a good measure of test progress and identification of failures for restricted setups, they might be insufficient for relaxed setups due to non-determinism.  
These findings are consistent with recent studies on misconceptions in DNN testing~\citep{ZohdinasabRGT23, RiccioT23}. Therefore, an interesting research direction is to develop metrics and evaluation methods that remain reliable in the presence of simulator non-determinism.

\section{Related Work}
\label{sec:relatedwork}
%https://arxiv.org/abs/2112.12331
 Recent research on flaky tests in software code-bases reveals their notable prevalence in both commercial and open-source contexts. Google reported that nearly 16\% of their 4.2 million test cases are flaky~\citep{google}, while 26\% of 3,871 distinct builds sampled from Microsoft's system failed due to flakiness~\citep{survey}.  The Microsoft Windows and Dynamics teams estimated a 5\% rate of flaky test failures~\citep{microsoft}, while the Randoop repository showed a similar rate of 5\% flaky tests for its open-source Java projects~\citep{Paydar2019AnES}. Our results in RQ1-1 reveal that the hard flaky test rates for the three ADS test setups adopted from the literature, namely \textsc{Pylot}~\citep{samota}, \textsc{Tran}~\citep{transfuser}, \textsc{Comp}~\citep{sbftgithub}, are 6\%, 32\%, and 1\% respectively for at least one of their fitness functions. Overall, between $4$\% and $68$\% of the generated tests across our five test setups exhibited noticeable variations in their fitness values. These results indicate that flakiness in ADS testing is comparable to flakiness in software code-bases.

Two recent simulation-based ADS testing studies have briefly noted the presence of flaky tests~\citep{arxivintro,salvo}. One study, based on the \textsc{Comp} setup, reported 1\% to 5\% flaky tests that were excluded from the results~\citep{arxivintro}.  The other study, based on  the SVL simulator~\citep{svl}, mitigated flakiness by voiding traffic lights~\citep{salvo}. We show that when test inputs and outputs are complex,  e.g., they include the elements in Figure~\ref{fig:concept}, flaky tests can be prevalent. In such situations, as shown in RQ1-3,  accounting for flakiness by rerunning tests significantly improves the results of randomized testing algorithms.

Detecting flaky tests often involves rerunning tests multiple times, which can be costly and time-consuming~\citep{survey}. Some methods approximate flakiness without requiring multiple test reruns by leveraging execution history and coverage information~\citep{deflaker,nondeterministic}. Some of these approaches rely on ML, NLP or probabilistic techniques to enhance their effectiveness in identifying flaky tests~\citep{FLASH,DBLP:conf/icse/Alshammari0H021,deflaker}. To our knowledge, no prior work has studied the presence, impact, or cost-effective prediction of flakiness in ADS testing. In RQ2-1 and RQ2-2, we demonstrate that ML classifiers can effectively identify flaky tests with a limited number of test reruns.

\section{Conclusion}
\label{sec:con}
This paper presents the first study evaluating the impact of flaky simulators on testing Autonomous Driving Systems (ADS). Our study includes combinations of two widely-used ADS simulators, CARLA and BeamNG, and three different ADS types. Our study shows that flakiness is  a common occurrence in ADS simulation-based testing.
We demonstrate that ML classifiers trained for each test setup are able to identify flaky ADS tests, requiring only a single run and achieving F1-scores of 85\%, 82\% and 96\% for three different ADS test setups. Considering the widespread occurrence of flaky tests in simulation-based ADS testing, it is crucial to evaluate the flaky test rates. If they prove to be significant, implementing mitigation strategies like test repetition or using ML classifiers, as demonstrated in this paper, can be beneficial. Alternatively, it is possible to assess test results using metrics that remain robust in the presence of ADS simulator flakiness.

Our study is unique in terms of the diversity of ADS test setups.  To the best of our knowledge, very few studies on ADS testing are performed on two or more simulators and three different ADS types. Nonetheless, further research is needed to understand the causes of flakiness in ADS simulators and to establish methods for assessing ADS testing algorithms given simulators' non-determinism.

\section{Data availability}
\label{sec:data}
Our online material include: (1)~a complete description and implementation of our test generators including test inputs, fitness functions and thresholds, (2)~scripts for reproducing our results, and (3)~raw datasets for the experiments~\citep{github}. Complementary experiment results, diagrams and statistical tests are  also included in the online material~\citep{sup_material}. 

%\input{tex/acknowledgments}
% The next two lines define the bibliography style to be used, and the bibliography file.
% \bibliographystyle{ACM-Reference-Format}

\section*{Acknowledgements}
We gratefully acknowledge the financial support received from NSERC of Canada through their Discovery program.

\section*{COI Statement}
The authors declared that they have no conflict of interest.

\balance
\bibliography{bib/ref}

\begin{thebibliography}{47}
\providecommand{\natexlab}[1]{#1}
\providecommand{\url}[1]{{#1}}
\providecommand{\urlprefix}{URL }
\expandafter\ifx\csname urlstyle\endcsname\relax
  \providecommand{\doi}[1]{DOI~\discretionary{}{}{}#1}\else
  \providecommand{\doi}{DOI~\discretionary{}{}{}\begingroup
  \urlstyle{rm}\Url}\fi
\providecommand{\eprint}[2][]{\url{#2}}

\bibitem[{uda(2016)}]{udacity:challenge}
 (2016) Udacity self-driving challenge 2.
  \url{https://github.com/udacity/self-driving-car/tree/master/challenges/challenge-2},
  accessed: 2019-10-11

\bibitem[{car(2022{\natexlab{a}})}]{carlafoundations}
 (2022{\natexlab{a}}) {Foundations}.
  \url{https://carla.readthedocs.io/en/latest/foundations/}, [Online; accessed
  15-November-2022]

\bibitem[{car(2022{\natexlab{b}})}]{carlaquickstart}
 (2022{\natexlab{b}}) Quick start.
  \url{https://carla.readthedocs.io/en/latest/start\_quickstart/}, [Online;
  accessed 15-November-2022]

\bibitem[{Raq(2022)}]{RaquelCompany}
 (2022) {R}aquel {U}rtasun’s tech company develops self-driving vehicle
  simulator.
  \url{https://www.thestar.com/business/2022/02/09/raquel-urtasuns-tech-company-develops-self-driving-vehicle-simulator.html},
  [Online; Accessed: May 2022]

\bibitem[{bea(2023)}]{beamng}
 (2023) {BeamNG.tech Website}. \url{https://beamng.tech}, [Online; accessed
  3-March-2023]

\bibitem[{car(2023)}]{carlatm}
 (2023) {Carla Challenge}.
  \url{https://carla.readthedocs.io/en/latest/adv\_traffic\_manager/}, [Online;
  accessed 1-February-2023]

\bibitem[{sbf(2023)}]{sbftgithub}
 (2023) Github repo for cyber-physical systems testing tool competition.
  \url{https://github.com/sbft-cps-tool-competition/cps-tool-competition},
  [Online; accessed 10-April-2023]

\bibitem[{svl(2023)}]{svl}
 (2023) Github repo for svl simulator: An autonomous vehicle simulator.
  \url{https://github.com/lgsvl/simulator}, [Online; accessed 10-April-2023]

\bibitem[{git(2023)}]{github}
 (2023) Github repo for the paper.
  \url{https://github.com/anonoymous9423013/anonymous_paper/}, [Online;
  accessed 10-April-2023]

\bibitem[{tra(2023)}]{transfuser}
 (2023) Github repo for transfuser: Imitation with transformer-based sensor
  fusion for autonomous driving.
  \url{https://github.com/autonomousvision/transfuser}, [Online; accessed
  10-April-2023]

\bibitem[{sup(2023)}]{sup_material}
 (2023) Online supplementary material for the paper.
  \url{https://github.com/anonoymous9423013/anonymous_paper/tree/main/supplementary_materials},
  [Online; accessed 26-April-2023]

\bibitem[{Abdessalem et~al.(2018)Abdessalem, Nejati, Briand, and
  Stifter}]{raja2018}
Abdessalem RB, Nejati S, Briand LC, Stifter T (2018) Testing vision-based
  control systems using learnable evolutionary algorithms. In: 2018 IEEE/ACM
  40th International Conference on Software Engineering (ICSE), IEEE, pp
  1016--1026

\bibitem[{Afzal et~al.(2021)Afzal, Katz, Le~Goues, and
  Timperley}]{afsoon-robotics}
Afzal A, Katz DS, Le~Goues C, Timperley CS (2021) Simulation for robotics test
  automation: Developer perspectives. In: 2021 14th IEEE Conference on Software
  Testing, Verification and Validation (ICST), pp 263--274

\bibitem[{Ahlgren et~al.(2021)Ahlgren, Bojarczuk, Drossopoulou, Dvortsova,
  George, Gucevska, Harman, Lomeli, Lucas, Meijer et~al.}]{ahlgren2021facebook}
Ahlgren J, Bojarczuk K, Drossopoulou S, Dvortsova I, George J, Gucevska N,
  Harman M, Lomeli M, Lucas SM, Meijer E, et~al. (2021) Facebook’s
  cyber--cyber and cyber--physical digital twins. In: Evaluation and Assessment
  in Software Engineering, pp 1--9

\bibitem[{Alshammari et~al.(2021)Alshammari, Morris, Hilton, and
  Bell}]{DBLP:conf/icse/Alshammari0H021}
Alshammari A, Morris C, Hilton M, Bell J (2021) Flakeflagger: Predicting
  flakiness without rerunning tests. In: 43rd {IEEE/ACM} International
  Conference on Software Engineering: Companion Proceedings, {ICSE} Companion
  2021, Madrid, Spain, May 25-28, 2021, {IEEE}, p 187

\bibitem[{Bell et~al.(2018)Bell, Legunsen, Hilton, Eloussi, Yung, and
  Marinov}]{deflaker}
Bell J, Legunsen O, Hilton M, Eloussi L, Yung T, Marinov D (2018) Deflaker:
  Automatically detecting flaky tests. In: 2018 IEEE/ACM 40th International
  Conference on Software Engineering (ICSE), pp 433--444

\bibitem[{Birchler et~al.(2023)Birchler, Khatiri, Bosshard, Gambi, and
  Panichella}]{arxivintro}
Birchler C, Khatiri S, Bosshard B, Gambi A, Panichella S (2023) Machine
  learning-based test selection for simulation-based testing of self-driving
  cars software. Empir Softw Eng 28(3):71

\bibitem[{Borg et~al.(2021)Borg, Abdessalem, Nejati, Jegeden, and
  Shin}]{BorgANJS21}
Borg M, Abdessalem RB, Nejati S, Jegeden F, Shin D (2021) Digital twins are not
  monozygotic - cross-replicating {ADAS} testing in two industry-grade
  automotive simulators. In: 14th {IEEE} Conference on Software Testing,
  Verification and Validation, {ICST} 2021, Porto de Galinhas, Brazil, April
  12-16, 2021, {IEEE}, pp 383--393

\bibitem[{Capon(1991)}]{capon:91}
Capon JA (1991) Elementary Statistics for the Social Sciences: Study Guide.
  Wadsworth Publishing Company, Belmont, CA, USA

\bibitem[{Chawla et~al.(2002)Chawla, Bowyer, Hall, and Kegelmeyer}]{smote}
Chawla NV, Bowyer KW, Hall LO, Kegelmeyer WP (2002) Smote: synthetic minority
  over-sampling technique. Journal of artificial intelligence research
  16:321--357

\bibitem[{Dosovitskiy et~al.(2017)Dosovitskiy, Ros, Codevilla, Lopez, and
  Koltun}]{carlapaper}
Dosovitskiy A, Ros G, Codevilla F, Lopez A, Koltun V (2017) {CARLA}: {An} open
  urban driving simulator. In: Proceedings of the 1st Annual Conference on
  Robot Learning, pp 1--16

\bibitem[{Dutta et~al.(2020)Dutta, Shi, Choudhary, Zhang, Jain, and
  Misailovic}]{FLASH}
Dutta S, Shi A, Choudhary R, Zhang Z, Jain A, Misailovic S (2020) Detecting
  flaky tests in probabilistic and machine learning applications. In:
  Proceedings of the 29th ACM SIGSOFT International Symposium on Software
  Testing and Analysis, Association for Computing Machinery, New York, NY, USA,
  ISSTA 2020, p 211–224, \doi{10.1145/3395363.3397366},
  \urlprefix\url{https://doi.org/10.1145/3395363.3397366}

\bibitem[{Gog et~al.(2021)Gog, Kalra, Schafhalter, Wright, Gonzalez, and
  Stoica}]{gog2021pylot}
Gog I, Kalra S, Schafhalter P, Wright MA, Gonzalez JE, Stoica I (2021) Pylot: A
  modular platform for exploring latency-accuracy tradeoffs in autonomous
  vehicles. In: 2021 IEEE International Conference on Robotics and Automation
  (ICRA), IEEE, pp 8806--8813

\bibitem[{Goutte and Gaussier(2005)}]{f1score}
Goutte C, Gaussier E (2005) A probabilistic interpretation of precision, recall
  and f-score, with implication for evaluation. In: Losada DE,
  Fern{\'a}ndez-Luna JM (eds) Advances in Information Retrieval, Springer
  Berlin Heidelberg, Berlin, Heidelberg, pp 345--359

\bibitem[{Hagan et~al.(1997)Hagan, Demuth, and Beale}]{hagan1997neural}
Hagan MT, Demuth HB, Beale M (1997) Neural network design. PWS Publishing Co.

\bibitem[{Haq et~al.(2020)Haq, Shin, Nejati, and
  Briand}]{DBLP:conf/icst/HaqSNB20}
Haq FU, Shin D, Nejati S, Briand LC (2020) Comparing offline and online testing
  of deep neural networks: An autonomous car case study. In: 13th {IEEE}
  International Conference on Software Testing, Validation and Verification,
  {ICST} 2020, Porto, Portugal, October 24-28, 2020, {IEEE}, pp 85--95

\bibitem[{Haq et~al.(2021)Haq, Shin, Nejati, and
  Briand}]{DBLP:journals/ese/HaqSNB21}
Haq FU, Shin D, Nejati S, Briand LC (2021) Can offline testing of deep neural
  networks replace their online testing? Empir Softw Eng 26(5):90

\bibitem[{Haq et~al.(2022)Haq, Shin, and Briand}]{samota}
Haq FU, Shin D, Briand L (2022) Efficient online testing for dnn-enabled
  systems using surrogate-assisted and many-objective optimization. In: 2022
  IEEE/ACM 44th International Conference on Software Engineering (ICSE), pp
  811--822, \doi{10.1145/3510003.3510188}

\bibitem[{Haq et~al.(2023)Haq, Shin, and Briand}]{haq2022manyobjective}
Haq FU, Shin D, Briand LC (2023) Many-objective reinforcement learning for
  online testing of dnn-enabled systems. In: 45th {IEEE/ACM} International
  Conference on Software Engineering, {ICSE} 2023, Melbourne, Australia, May
  14-20, 2023, {IEEE}, pp 1814--1826

\bibitem[{Harman and McMinn(2010)}]{5342440}
Harman M, McMinn P (2010) A theoretical and empirical study of search-based
  testing: Local, global, and hybrid search. IEEE Transactions on Software
  Engineering 36(2):226--247, \doi{10.1109/TSE.2009.71}

\bibitem[{Herzig and Nagappan(2015)}]{microsoft}
Herzig K, Nagappan N (2015) Empirically detecting false test alarms using
  association rules. In: 2015 IEEE/ACM 37th IEEE International Conference on
  Software Engineering, vol~2, pp 39--48

\bibitem[{Luke(2013)}]{metaheuristicsbook}
Luke S (2013) Essentials of Metaheuristics, 2nd edn. Lulu, available for free
  at http://cs.gmu.edu/$\sim$sean/book/metaheuristics/

\bibitem[{Luo et~al.(2014)Luo, Hariri, Eloussi, and
  Marinov}]{DBLP:conf/sigsoft/LuoHEM14}
Luo Q, Hariri F, Eloussi L, Marinov D (2014) An empirical analysis of flaky
  tests. In: Cheung S, Orso A, Storey MD (eds) Proceedings of the 22nd {ACM}
  {SIGSOFT} International Symposium on Foundations of Software Engineering,
  (FSE-22), Hong Kong, China, November 16 - 22, 2014, {ACM}, pp 643--653

\bibitem[{Matinnejad et~al.(2017)Matinnejad, Nejati, and
  Briand}]{MatinnejadNB17}
Matinnejad R, Nejati S, Briand LC (2017) Automated testing of hybrid
  simulink/stateflow controllers: industrial case studies. In: Bodden E,
  Sch{\"{a}}fer W, van Deursen A, Zisman A (eds) Proceedings of the 2017 11th
  Joint Meeting on Foundations of Software Engineering, {ESEC/FSE} 2017,
  Paderborn, Germany, September 4-8, 2017, {ACM}, pp 938--943

\bibitem[{Micco(2018)}]{google}
Micco J (2018) Advances in continuous integration testing at google

\bibitem[{Nguyen et~al.(2021)Nguyen, Huber, and Gambi}]{salvo}
Nguyen V, Huber S, Gambi A (2021) Salvo: Automated generation of diversified
  tests for self-driving cars from existing maps. In: 2021 IEEE International
  Conference on Artificial Intelligence Testing (AITest), pp 128--135

\bibitem[{Parry et~al.(2021)Parry, Kapfhammer, Hilton, and McMinn}]{survey}
Parry O, Kapfhammer GM, Hilton M, McMinn P (2021) A survey of flaky tests. ACM
  Trans Softw Eng Methodol 31(1), \doi{10.1145/3476105},
  \urlprefix\url{https://doi.org/10.1145/3476105}

\bibitem[{Paydar and Azamnouri(2019)}]{Paydar2019AnES}
Paydar S, Azamnouri A (2019) An experimental study on flakiness and fragility
  of randoop regression test suites. In: Fundamentals of Software Engineering

\bibitem[{Riccio and Tonella(2023)}]{RiccioT23}
Riccio V, Tonella P (2023) When and why test generators for deep learning
  produce invalid inputs: an empirical study. In: 45th {IEEE/ACM} International
  Conference on Software Engineering, {ICSE} 2023, Melbourne, Australia, May
  14-20, 2023, {IEEE}, pp 1161--1173

\bibitem[{Samak et~al.(2020)Samak, Samak, and Kandhasamy}]{pidpaper}
Samak CV, Samak TV, Kandhasamy S (2020) Control strategies for autonomous
  vehicles. CoRR abs/2011.08729,
  \urlprefix\url{https://arxiv.org/abs/2011.08729}, \eprint{2011.08729}

\bibitem[{Shi et~al.(2016)Shi, Gyori, Legunsen, and Marinov}]{nondeterministic}
Shi A, Gyori A, Legunsen O, Marinov D (2016) Detecting assumptions on
  deterministic implementations of non-deterministic specifications. In: 2016
  IEEE International Conference on Software Testing, Verification and
  Validation (ICST), pp 80--90

\bibitem[{Ulbrich et~al.(2015)Ulbrich, Menzel, Reschka, Schuldt, and
  Maurer}]{definitionpaper}
Ulbrich S, Menzel T, Reschka A, Schuldt F, Maurer M (2015) Defining and
  substantiating the terms scene, situation, and scenario for automated
  driving. In: 2015 IEEE 18th International Conference on Intelligent
  Transportation Systems, pp 982--988, \doi{10.1109/ITSC.2015.164}

\bibitem[{Vargha and Delaney(2000)}]{vargha}
Vargha A, Delaney HD (2000) A critique and improvement of the cl common
  language effect size statistics of mcgraw and wong. Journal of Educational
  and Behavioral Statistics 25(2):101--132

\bibitem[{Witten et~al.(2011)Witten, Frank, and Hall}]{dataminingbook}
Witten IH, Frank E, Hall MA (2011) Data Mining: Practical Machine Learning
  Tools and Techniques, 3rd edn. Morgan Kaufmann Series in Data Management
  Systems, Morgan Kaufmann, Amsterdam

\bibitem[{Zeller et~al.(2023)Zeller, Gopinath, B{\"o}hme, Fraser, and
  Holler}]{fuzzingbook2023:Coverage}
Zeller A, Gopinath R, B{\"o}hme M, Fraser G, Holler C (2023) Code coverage. In:
  The Fuzzing Book, CISPA Helmholtz Center for Information Security,
  \urlprefix\url{https://www.fuzzingbook.org/html/Coverage.html}, retrieved
  2023-01-07 13:54:15+01:00

\bibitem[{Zhong et~al.(2023)Zhong, Kaiser, and Ray}]{zhong}
Zhong Z, Kaiser G, Ray B (2023) Neural network guided evolutionary fuzzing for
  finding traffic violations of autonomous vehicles. IEEE Transactions on
  Software Engineering 49(4):1860--1875, \doi{10.1109/TSE.2022.3195640}

\bibitem[{Zohdinasab et~al.(2023)Zohdinasab, Riccio, Gambi, and
  Tonella}]{ZohdinasabRGT23}
Zohdinasab T, Riccio V, Gambi A, Tonella P (2023) Deephyperion: Exploring the
  feature space of deep learning-based systems through illumination search. In:
  Engels G, Hebig R, Tichy M (eds) Software Engineering 2023, Fachtagung des
  GI-Fachbereichs Softwaretechnik, 20.-24. Februar 2023, Paderborn,
  Gesellschaft f{\"{u}}r Informatik e.V., {LNI}, vol {P-332}, pp 131--132

\end{thebibliography}

% If your work has an appendix, this is the place to put it.
% \appendix

\end{document}